\documentclass[a4paper,fleqn,usenatbib,apj]{emulateapj}

\usepackage[T1]{fontenc}
\usepackage{ae,aecompl}
\usepackage{graphicx}
\usepackage{amsmath}
\usepackage{amssymb}
\usepackage{calligra}
\usepackage{dcolumn}   
\usepackage{bm}        
\usepackage{amsfonts,amsmath,amssymb,mathrsfs}
\usepackage{color}
\usepackage[table]{xcolor} 
\usepackage{hyperref}
\hypersetup{
  colorlinks=true,        
  linkcolor=blue,         
  citecolor=cyan,         
}
\usepackage{natbib,times}
\newcommand{\ie}{i.e.,~}
\newcommand{\eg}{e.g.,~}

\begin{document}

\title[Relativistic Common Envelope]{Common-envelope Dynamics of a
  Stellar-mass Black Hole: General Relativistic Simulations}
\bigskip

\author{A. Cruz-Osorio\altaffilmark{1}, L. Rezzolla \altaffilmark{1,2}}
\altaffiltext{1}{Institute for Theoretical Physics,
  Frankfurt, Ruth-Moufang-Stra{\ss}e 1, D-60438 Frankfurt am Main,
  Germany.}
\altaffiltext{2}{School of Mathematics, Trinity College, Dublin 2,
  Ireland}

\begin{abstract}
With the goal of providing more accurate and realistic estimates of the
secular behavior of the mass accretion and drag rates in the
``common-envelope'' scenario encountered when a black hole or a neutron
star moves in the stellar envelope of a red supergiant star, we have
carried out the first general relativistic simulations of the accretion
flow onto a nonrotating black hole moving supersonically in a medium with
regular but different density gradients.
The simulations reveal that the supersonic motion always rapidly reaches
a stationary state and it produces a shock cone in the downstream part of
the flow. In the absence of density gradients we recover the
phenomenology already observed in the well-known Bondi--Hoyle--Lyttleton
accretion problem, with super-Eddington mass accretion rate and a shock
cone whose axis is stably aligned with the direction of motion. However,
as the density gradient is made stronger, the accretion rate also
increases and the shock cone is progressively and stably dragged toward
the direction of motion. With sufficiently large gradients, the
shock-cone axis can become orthogonal to the direction, or even move in
the upstream region of the flow in the case of the largest density
gradient.
Together with the phenomenological aspects of the accretion flow, we have
also quantified the rates of accretion of mass and momentum onto the
black hole. Simple analytic expressions have been found for the rates of
accretion of mass, momentum, drag force, and bremsstrahlung luminosity,
all of which have been employed in the astrophysical modelling of the secular
evolution of a binary system experiencing a common-envelope evolution.
We have also compared our results with those of previous studies in
Newtonian gravity, finding similar phenomenology and rates for motion in
a uniform medium. However, differences develop for nonzero density
gradients, with the general relativistic rates increasing almost
exponentially with the density gradients, while the opposite is true for
the Newtonian rates. Finally, the evidence that mass accretion rates well
above the Eddington limit can be achieved in the presence of nonuniform
media, increases the chances of observing this process also in binary
systems of stellar-mass black holes.
\end{abstract}

\maketitle

\section{Introduction}
\label{sec:intro}

The gravitational-wave detections recently made by the LIGO and Virgo
Collaborations have provided the long-sought observational
confirmation of the existence of binary black hole systems
\citep{Abbot2016-GW-detection-prl, Abbot2016g, Abbott2017a,
  Abbott2017g}. In addition, and more recently, the gravitational-wave
emission from a merging binary system of neutron stars has been detected
for the first time \citep[GW170817;][]{Abbott2017}, confirming many of the
theoretical expectations behind the properties of these mergers
\citep[see][for recent reviews]{Baiotti2016, Paschalidis2017}. In
addition, great expectations are in place that a binary system comprising
a black hole and a neutron star \citep{ShibataTaniguchilrr-2011-6} will
be detected as the detectors resume data collection after their upgrades.

The long history of all of these compact-object binaries is characterized
by a stage, the "common-envelope'' evolution, which has been
the focus of a lot of attention in the more remote and recent past
\citep[see, e.g.,][]{Livio1988, Taam2000, Taam2010,
  Ivanova2013, MacLeod2015, Murguia-Berthier2017845}. Despite the large
bulk of work made to describe this phase of the evolution of the binary
system, the quantitative aspects of the secular evolution are far from
being clear.

\begin{figure}
  \center
  \includegraphics[width=1.0\columnwidth]{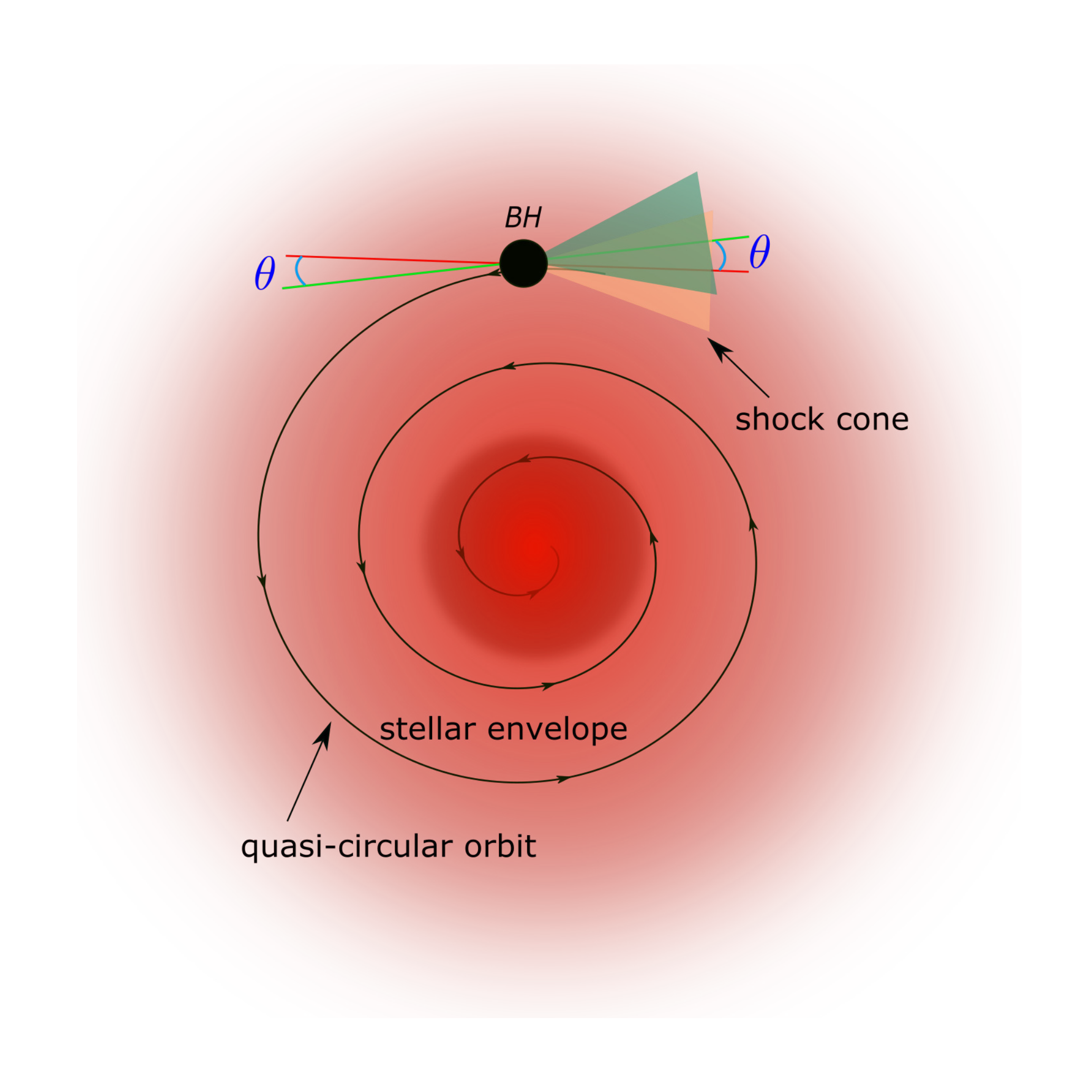}
  \caption{Schematic diagram -- not to scale -- of the system being
    simulated: a black hole moves supersonically in the envelope of a
    supergiant red star, producing a shock cone in the downstream region
    of the flow. In doing so, it encounters a density gradient described
    with the dimensionless parameter $\epsilon_{\rho}$. The modifications
    of the trajectory can be measured through the deviation of the angle
    of the shock cone with respect to the upstream flow.}
  \label{fig:tzo}
\end{figure}

Among the aspects that are clear of this picture is that common-envelope
evolution could involve two different scenarios of compact binaries. The
first one comprises a red giant or supergiant star of mass $\simeq
15-20\,M_{\odot}$ and radius of $\sim 1000\,R_{\odot}$ containing at its
core a neutron star with mass $1.3-2.2\,M_{\odot}$ \citep{Margalit2017,
  Shibata2017c,Rezzolla2017, Ruiz2017} and radius $\sim 10\,{\rm km}$
\citep{Annala2017, Most2018}; this is also known as a "Thorne--Zytkow
object'' \citep{Thorne75, Hutilukejiang2018}, where the astronomical
object HV-2112 represents a possible candidate \citep{Levesque2014,
  Maccarone2016}. The second possibility is instead represented by a
binary where the compact object is replaced by a stellar-mass black hole
as illustrated in the cartoon in Fig. \ref{fig:tzo}.

In both cases, the lifetime of the common-envelope phase is rather brief
and less than {$300$ yr}, during which the orbital separation between
the two components of the binary reduces by factor of $\sim 100$. This
result was first pointed out in the pioneering work of \citet{Sparks74}
and \citet{Paczynski76}, who showed through analytical estimates that the
common-envelope phase is responsible for the reduction of the orbital
period of the binary system via the loss of mass and orbital angular
momentum in the binary, which transforms the orbital energy of the binary
into kinetic energy of the matter in the envelope, {which is heated and
  gains angular momentum} \citep{Paczynski76, Livio1988, Taam2010}; this
reduction of the orbit can then lead to the merger of the binary system
or to the production of a very tight binary
\citep{Ivanova2013}. Furthermore, the common-envelope dynamics is thought
to be responsible for some of the phenomenology observed in X-ray
binaries, close binary systems, and progenitors of gamma-ray bursts and
even of Type IIP supernovae, as suggested by
\citet{Ivanova201339}.

Given the difficulties in modeling the nonlinear processes that take
place during the common-envelope evolution, over the years several
  numerical simulations have been performed, either in one spatial
  dimension  \citep[1D;][]{Taam78, Meyer79, Delgado1980,
  Podsiadlowski2001, Ivanova2015, Clayton2017}, {in two dimensions}
\citep[2D;][]{Bodenheimer1984, Armitage2000, Blondin2009, Blondin2013}, {and in
  three dimensions} \citep[3D;][]{Ricker2012, Nandez2015, MacLeod2015, Ohlmann2016, 
  Staff2016, Nandez2016, Iaconi2017, Murguia-Berthier2017845}. Overall, {this bulk}
of rather diverse simulations has been able to cover binary systems where
the compact object is either a white dwarf, a neutron star or a black
hole \citep[see, e.g.,][]{MacLeod2015}.

A common feature of all of these simulations is that they have been
performed within a Newtonian description of gravity. While this
may be a reasonable approximation near the central regions of the red
supergiant and possibly in the case of a white dwarf, it certainly ceases
to be so near the compact object, be it a neutron star or a black hole,
and where general relativistic corrections can be
considerable. {We here present results} from the first
general relativistic simulations of the common-envelope evolution
of a system composed of a red supergiant star interacting with a
stellar-mass Schwarzschild black hole; such a system could represent, for
instance, the progenitor of the X-ray binary system Cygnus X-3
\citep{Postnov2014}.

For simplicity, and as a first step in a series of investigations of this
process, we here consider the black hole to be a nonrotating (\ie a
Schwarzschild) solution and carry out our simulations in a 2D cut, which is
representative of the accretion flow on the equatorial plane. {In doing
  so, we consider the matter in the flow to be a "test matter", hence
  following the background spacetime geometry of the black hole only; in
  this way we explicitly neglect the curvature induced by the supergiant
  stellar core}\footnote{{In this respect, 3D Newtonian simulations have
    the advantage of being able to consistently include the
    core-companion gravity and provide a more consistent evolution of the
    density profile as the black hole inspirals toward the stellar core
    \citep{Ivanova2013}.}}. {We also} provide analytic expressions for
the mass and angular momentum accretion rates and for the angle of the
deviation of the shock cone due to drag forces as a function of the density
gradient and other reference quantities of the system. Such expressions
can represent useful inputs in the phenomenological modeling of the
common-envelope evolution.

The plan of the paper is as follows: In Sec. \ref{sec:Num} we describe
the metric of Schwarzschild spacetime in spherical coordinates written in
3+1 decomposition, the relativistic hydrodynamic equations, and the
computational infrastructure employed in their numerical solution via the
\texttt{CAFE} code. We also describe the characteristics of the
common-envelope scenario we are investigating, the construction of the
initial data configurations, and the boundary conditions adopted to
guarantee a stationary solution. In Sec. \ref{sec:Res} we report our
results showing the morphology of the fluid dynamics, the rates of
accretion of rest mass and angular momentum, and other related
quantities, such as the angle of deviation of the shock cone and the
produced bremsstrahlung luminosity. In Sec. \ref{sec:Sum} we summarize
our conclusions and the prospects for future work. Two appendices
  complement the main text and provide supplemental information. In
  Appendix \ref{sec:appA}, we calculate the inspiral timescales as
  derived from our measured accretion rates and explore how the accretion
  rates are modified by a different prescription for the initial data. In
  Appendix \ref{sec:appB}, on the other hand, we present convergence
  tests and a comparison of numerical outcomes from two numerical codes.
Hereafter, we will adopt the Einstein convention on sums over repeated
indices and use geometrized units where $G=c=1$.

\section{Mathematical and numerical setup}
\label{sec:Num}

\subsection{General relativistic hydrodynamics}
\label{sec:Euler}

The simulations carried out involve the solution of the equations of
relativistic hydrodynamics in which the matter is considered to behave as
a test fluid in the curved spacetime of the astrophysical compact object
and thus not to affect it \citep{Rezzolla_book:2013}. For simplicity, and
to build up our physical understanding of the relativistic corrections
encountered when studying the evolution of the common-envelope phase in a
general relativistic framework, we here consider the background spacetime
to be that of a nonrotating black hole of mass $M$, described therefore
by the Schwarzschild metric. For the latter and for its convenience in
numerical implementation, we use horizon-penetrating
Eddington--Finkelstein coordinates in a 3+1 decomposition where the line
element of the spacetime is expressed as
\begin{equation}
ds^{2}= -(\alpha^{2} - \beta_{i}\beta^{i})dt^{2} + 2\beta_{i}dx^{i}dt +
\gamma_{ij}dx^{i}dx^{j}\,,
\label{eq:metric}
\end{equation}
where $\beta^i$ is the shift vector, $\alpha$ is the lapse function, and
$\gamma_{ij}$ are the components of the spatial three-metric
\citep{Alcubierre:2008,Rezzolla_book:2013}. More specifically, the
explicit expressions for these functions are given by
\begin{eqnarray}
\label{eq:schw_1}
\alpha &=& \left( 1+\frac{2M}{r}\right)^{-1/2}\,, \\
\beta^i &=& \left(
\frac{2M}{r(1+ 2M/r)} , 0, 0 \right)\,,
\end{eqnarray}
and
\begin{eqnarray}
\label{eq:schw_2}
\gamma_{ij} =\left( 
	\begin{array}{ccc}
1+{2M}/{r} & \ \ \ 0    & \ \ \ 0 \\
0          & \ \ \ r^2  & \ \ \ 0\\
0          & \ \ \ 0    & \ \ \ r^2 \sin^2\theta 
         \end{array}
\right)\,.
\end{eqnarray}

Given a generic curved spacetime with associated four-metric
$g_{\mu\nu}$, \eg  equations \eqref{eq:schw_1}--\eqref{eq:schw_2}, and with
associated covariant derivative $\nabla_{\mu}$, the equations of
relativistic hydrodynamics can be written as simple conservation laws of
the energy-momentum tensor $T^{\mu \nu}$ and of the rest-mass current
$J^{\mu}$ \citep{Rezzolla_book:2013},
\begin{eqnarray}
  \nabla_{\mu}(T^{\mu\nu}) &=& 0\,,\label{eq:SEcons}\\
  \nabla_{\mu}(J^{\mu}) = \nabla_{\mu}(\rho u^{\mu}) &=& 0\,.
  \label{eq:masscons}
\end{eqnarray}

Assuming the fluid to be a perfect one, the corresponding energy-momentum
tensor is given by
\begin{eqnarray}
  T^{\mu \nu} =\rho h u^{\mu}u^{\nu} + pg^{\mu\nu}\,,
  \label{eq:stress-energy}
\end{eqnarray}
where $\rho$ is the rest-mass density of a fluid element; $u^{\mu}$ is the
four-velocity of the fluid; $h=1+\epsilon+p/\rho$ is the specific
enthalpy, with $\epsilon$ the specific internal energy; and $p$ the
pressure.

The set of relativistic hydrodynamic equations is then closed by an
equation of state relating the pressure to other thermodynamics
quantities in the fluid. As customary in the calculations of this type,
we here assume the equation of state to be that of an ideal fluid, where
\begin{equation}
  p = \rho \epsilon  (\Gamma-1) \,,
  \label{eq:EOS}
\end{equation}
where $\Gamma$ is the adiabatic index, for which we explore the two
limits of $\Gamma= 5/3$ and $\Gamma=4/3$, corresponding to the properties
of a cold degenerate electron fluid and a completely degenerate
ultrarelativistic electron fluid, respectively
\citep{Rezzolla_book:2013}.

\subsection{Numerical Methods}
\label{sec:Met}

For the numerical solution of relativistic hydrodynamic equations we
employ a conservative formulation, also known as the Valencia formulation
\citep{Banyuls97}, through which a number of important mathematical
properties of the set of hyperbolic equations are preserved
\citep{Rezzolla_book:2013}. In our 2D system of spherical {azimuthal}
coordinates representing the orbital (or equatorial) plane of the
common-envelope scenario, the relativistic hydrodynamic equations in the
Valencia formulation take the form
\begin{eqnarray}
\partial_{t} \boldsymbol{U} + \partial_{r} \boldsymbol{F}^{r}
+\partial_{\phi} \boldsymbol{F}^{\phi} = \boldsymbol{S} -
\frac{1}{2}\partial_{r} \log (\gamma)\boldsymbol{F}^{r}\,,
\label{eq:flux_conservative}
\end{eqnarray}
where $\gamma=\det(\gamma_{ij})$ is the determinant of the spatial
three-metric and $\boldsymbol{U} :=(D, S_{r}, S_{\phi}, \tau)$ is the vector
of the ``conserved'' variables
\begin{eqnarray}
\boldsymbol{U} &:=& (D, S_{r}, S_{\phi}, \tau) \nonumber\\
&=& \left( \rho W , \rho h W^{2}v_{r}, \rho h W^{2}v_{\phi}, \rho h W^{2}  -p- D \right) \,, 
\end{eqnarray}
which depend on the primitive variables with vector $\boldsymbol{P} =
(\rho, v^{i}, p, \epsilon)$. The vectors $\boldsymbol{F}^{r}$ and
$\boldsymbol{F}^{\phi}$ contain instead the ``fluxes'' along the $r$ and
$\phi$ spatial coordinates
\begin{eqnarray}
\boldsymbol{F}^{r}& =& \alpha \left( Dv^{r}, S_{r}v^{r} +  p\,, S_{\phi}v^{r}, (\tau + p)v^{r}  \right)\,, \\
\boldsymbol{F}^{\phi} &=& \alpha\left( D v^{\phi}, S_{r}v^{\phi},  S_{\phi}v^{\phi} +  p, ( \tau + p)v^{\phi} \right)\,, 
\end{eqnarray}
while $\boldsymbol{S}$ is the source vector and has explicit components
given by
\begin{eqnarray}
\boldsymbol{S} &=&  \alpha  \left(0, \frac{1}{2} T^{\mu \nu}
\partial_{r} g_{\mu \nu}, \frac{1}{2} T^{\mu \nu}  \partial_{\phi} g_{\mu
  \nu}, T^{\mu t}\partial_{\mu}\alpha
- T^{\mu \nu} \Gamma^{t}_{\mu\nu} \alpha \right) \,.
\nonumber\\
\end{eqnarray}
Here $\Gamma^{\alpha}_{\mu \mu} $ are the Christoffel symbols, $W
:=\alpha u^{0}$ is the Lorentz factor, where $v^i$ and
$v_i=\gamma_{ij}v^j$ are the components of the spatial three-velocity of
the fluid measured by a (normal) Eulerian observer. These
three-velocities are related to the spatial components of the
four-velocity by $v^{i}=u^{i}/W + \beta^{i}/\alpha$.

The relativistic hydrodynamic equations are solved numerically using
high-resolution shock-capturing methods \citep[HRSC;][]{Rezzolla_book:2013}
and, in particular, the Harten, Lax, van Leer, and Einfeldt (HLLE)
approximate Riemann solver \citep{Harten83, Einfeldt1991} combined with
the "minmod'' second-order total variation diminishing (TVD)
reconstruction approach at cell interfaces \citep[see][for more
  details]{Keppens2012, Porth2017}. The time evolution of the set of
partial differential equations is realized via a method-of-lines approach
and a third-order Runge-Kutta scheme, which guarantees the TVD property
\citep{Shu88}.

As mentioned above, the numerical code employed for the numerical
solution is the \texttt{CAFE} code, which has been developed to solve the
equations of relativistic hydrodynamics in 2D slab symmetry
\citep{Cruz2012, Lora2015219,Cruz2016}, 2D axial symmetry
\citep{Lora2013}, and magnetohydrodynamics (MHD) in 3D
\citep{Lora2015}. The numerical grid employs radial and azimuthal
coordinates and is uniformly spaced with $N_r \times N_{\phi} = 2000
\times 256$ cells. {The radial grid, in particular, extends from $r_{\rm
    exc}$ to $r_{\rm max}$, where $r_{\rm exc}$ is the excision radius
  that is located inside the event horizon, \ie at $r_{\rm exc}=1.5\, M\,
  (\sim 8.9\, {\rm km})$ and $r_{\rm max}=10\, r_{\rm acc}\, \sim (1190\,
  {\rm km})$ is expressed in terms of the accretion radius, $r_{\rm
    acc}$, whose definition is discussed below; the angular coordinate,
  $\phi$, on the other hand, spans the full internal $\phi \in [0 , 2 \pi
  ]$. The effective resolution $(\Delta r, \Delta \phi)= (0.1 \, M,
  0.0386 \, {\rm rad})$ in geometrized units, or, equivalently, $\Delta r
  \sim 0.6 \, {\rm km}$.}  Finally, the timestep size is constant and
satisfies the Courant-Friedrich-Lewy (CFL) stability condition, with
$\Delta t = \tfrac{1}{4}\, \min(\Delta r, \Delta \phi)$.

\subsection{Common-envelope setup}

In our modeling, we assume that the common-envelope phase takes place in
a binary system composed of a stellar-mass black hole with mass $M$ and a
more massive red supergiant with mass $M_{\rm star}$ and radius
  $R_{\rm star}$. This scenario could occur either after the collapse of
a neutron star to a black hole in a binary system initially containing a
red supergiant and a neutron star or in a binary system with a red
supergiant and a massive star that collapse directly to a black hole
\citep{Postnov2014, MacLeod2015}.
Assuming furthermore that the binary is in a quasi-circular orbit with
radius $R$, the black hole will have a Newtonian linear velocity that can
be easily estimated from the Keplerian motion to be $v_{_{\rm CE}} =
(m(R) + M)/R$, where $m(R)$ is the mass enclosed in a sphere with radius
$R$, the latter being necessarily $R<R_{\rm star}$ within the
common-envelope phase. Because it is numerically convenient to place the
black hole at the center of our coordinate system and to consider
therefore the relative motion of the stellar envelope, we can reverse the
reference systems and thus take the asymptotic velocity of the fluid
$v_{\infty}$ as $v_{_{\rm CE}} = v_{\infty}$. We note that, strictly
speaking, the Keplerian velocity at these separations is
$\mathcal{O}(10^2-10^3) \,{\rm km/s}$ giving velocities
$\mathcal{O}(10^{-3}-10^{-4}) \,{\rm c}$, in geometrical units and thus
too small to be used in numerical simulations\footnote{A small
    initial asymptotic velocity will lead to a very large accretion
    radius (see Equation \eqref{eq:racc} for a definition) and hence to even
    larger computational domains since the latter have to cover
    length scales that are tens of accretion radii.}. However, as long as
the proper Mach number is chosen for the simulations (\ie
$\mathcal{M}=1-5$ for typical common-envelope scenarios;
\citealt{MacLeod2017}), it is possible to perform simulations with
effectively larger velocities (\ie $\mathcal{O}(10^4)\, {\rm km/s}$ in
our case) and then rescale all the results to smaller initial velocities
through scaling relations (see Equations \eqref{eq:fitmrate} and
  \eqref{eq:fitmom} and discussion in Sec. \ref{sec:ma_dr} and Appendix
  \ref{sec:iotid}). A very similar approach has already been employed by
\citet{MacLeod2015} and \citet{MacLeod2017}, who have used $v_{\infty}=1$
in a fully Newtonian context.

Also for the rest-mass density distribution, we make the rather
  simplified but reasonable assumption that it is given by a
  constant-density core of density $\bar{\rho}$ and that then falls
  exponentially up to the surface. Assuming a coordinate system with
  origin in the center on the star, we thus express the rest-mass density
  as\footnote{Note that the density profile given by expression
  \eqref{eq:dence} is the result of the integration of
  Eq. \eqref{eq:gradient} defined below \citep[see][for more
    details]{MacLeod2015}.}
\begin{equation}
\label{eq:dence}
\rho(r)=
\begin{cases}
  \bar{\rho}={\rm const.} \quad &{\rm for}\quad 0 < r < \bar{r}  \\
  \rho_{0} \exp{\left[-\epsilon_{\rho}(r-r_{0})/r_{\rm acc} \right]}\,,
  \quad &{\rm for}\quad \bar{r} \leq r \leq R_{\rm star}
  \end{cases}
\end{equation}
where $r_0 > \bar{r}$ is the radial position of the black hole within the
common envelope (see below) and $\rho_{0}$ is the rest-mass density at
that position (\ie $\rho_{0}:=\rho(r=r_0)$); clearly, from expression
\eqref{eq:dence} it follows that $\bar{\rho}:=\rho(r=\bar{r})>
\rho_0$. Note also that we have introduced the "accretion radius'',
$r_{\rm acc}$, which represents the length scale over which the
gravitational effects of the compact object dominate over the dynamics of
the fluid, and which naturally appears when considering the problem of
accretion onto a compact object moving in a medium, \ie a
Bondi--Hoyle--Lyttleton accretion \citep{Hoyle1939, Bondi1944,
  Rezzolla_book:2013}. Its definition is therefore
\begin{equation}
  \label{eq:racc}
  r_{\rm acc} := \frac{M}{c^{2}_{s, \infty} + v^{2}_{\infty}}=
        \frac{M}{c^{2}_{s, \infty} (1+ {\cal M}^2_{\infty})}\,,
\end{equation}
and thus in terms of the asymptotic value of Newtonian Mach number ${\cal
  M}_{\infty}:=v_{\infty}/c_{s, \infty}$, where $c_{s, \infty}$ is the
asymptotic sound speed. As will become clearer when we introduce the
reference mass accretion rate (see Equation \eqref{eq:canmdot}), the accretion
radius \eqref{eq:racc} plays a fundamental role in establishing what is
the amount of matter that is accreted onto the black hole. More
importantly, the accretion radius (and hence the mass accretion rate) 
depends not only on the asymptotic velocity $v_{\infty}$, but also on the
asymptotic sound speed $c_{s, \infty}$.

Arguably, among the most important quantity in the description of the
common-envelope evolution is the strength of the density gradient that
the compact object experiences as it moves in the stellar envelope. A
convenient way to express this contrast is to normalize the rest-mass
density scale height
\begin{equation}
\mathcal{L}_{\rho} := - \frac{\rho}{d\rho/dr} \,,
\end{equation}
with the other characteristic length scale of the problem, namely, the
accretion radius, to obtain the ``dimensionless accretion radius''
\begin{equation}
  \epsilon_{\rho} := - r_{\rm acc} \frac{d\rho/dr}{\rho}
    = \frac{r_{\rm acc}}{\mathcal{L}_{\rho}}
   \,.
   \label{eq:gradient}
\end{equation}
In practice, $\epsilon_{\rho}$ is here treated as a free parameter, and in the
simulations we have explored various values, which are summarized in
Tab. \ref{tab:params}.

\subsection{Initial and boundary conditions}

For the initial stellar model we follow the prescription presented by
\citet{MacLeod2015}, thus assuming a binary system composed of a red
supergiant star with mass $M_{\rm star} = 16\,M_{\odot}$ and a
stellar-mass black hole with mass $M=4\, M_{\odot}$, so that the
corresponding mass ratio in the binary is $q := M/M_{\rm star}=0.25$. The
black hole is placed at $r_{0}=R_{\rm star}/2 \simeq
  400\,R_{\odot}\simeq 2.78 \times 10^{8} \,{\rm km}$ and the rest-mass
density there is assumed to be $\rho(r_0) =: \rho_0 = \rho_{\infty}
\simeq 9.51 \times 10^{-9}\, {\rm g/cm}^{3}$. With such a value
  and an intermediate value for the density gradient
  $\epsilon_{\rho}=0.5$, the integration of the rest-mass density
  \eqref{eq:dence} yields a supergiant star with 16 $M_{\odot}$.

For such a system, the merger timescale is given by \citep{Faber2012:lrr}
\begin{equation}
  \tau_{\rm merg}= 2.2\times 10^{8} \frac{1}{4q(1+q)} \left(
  \frac{a}{R_{\odot}}\right)^{4} \left( \frac{M_{\rm star}}{1.4\,
    M_{\odot}}\right)^{-3}\ {\rm yr}\,,
\end{equation}
and because our simulations are carried out over a timescale of
$\tau_{\rm sim} \simeq 20000\, M_\odot \sim 0.1\,{\rm sec}$, it is
perfectly reasonable to assume that the density profile does not to
change over the time of the simulations.

In the reference frame where the black hole is at rest, the
common-envelope material is seen moving moving at a constant velocity that
is uniform within the computational domain and set assuming that  the sound
speed of the fluid is $c_{s, \infty}=0.1$ and that the fluid is
supersonic with Mach number ${\cal M}_{\infty}=2$ (as a result, the
asymptotic fluid velocity is $v_{\infty}=0.2$).

After mapping the spherical polar coordinate system $(r,\phi)$ to
  a Cartesian $(x,y)$ system where the stellar center is at the position
  $(x_0, y_0)$ \citep[see][for more details on the mapping]{Cruz2012},
  the common envelope is taken to be moving in the positive
  $x$-direction, \ie $v^{x}=v_{\infty}$, and the matter is taken to have a
  rest-mass gradient \emph{only} in the $y$-direction (see 
    Equation \eqref{eq:dence})\footnote{A black hole moving along a
    quasi-circular orbit across a density distribution that is
    spherically symmetric with respect to the center of the orbit will
    experience a density gradient only in the direction orthogonal to the
    direction of motion. Given the small size of the region we are
    simulating, the circular motion is essentially linear and taken to be
    along the $x$-direction, so that the gradient is only in the
    $y$-direction.}. As a result, the initial rest-mass density profile
  is set to be $\rho_{\rm in} := \rho(y) = \rho_{0} \exp{\left[
      -\epsilon_{\rho} y/r_{\rm acc} \right]}$, where $y\in [-\bar{r},
    \bar{r}]$, and we set the edge of our numerical grid to coincide with
  the edge of the constant-density stellar core, \ie $r_{\rm
    max}=\bar{r}=10\,r_{\rm acc}$. The resulting accretion radius after
  substituting the asymptotic sound speed and the asymptotic velocity in
  Eq. \eqref{eq:racc} is $r_{\rm acc}=20 \, M \sim 119 \, {\rm km}$.

Once a choice is made for the initial rest-mass density distribution and
the equation of state has been fixed to be that of an ideal fluid, the
initial pressure is determined by the definition of the sound speed as
\begin{equation}
p_{\rm in}=c^{2}_{s, \infty} \frac{\Gamma -1}{\Gamma(\Gamma -1)
  -c^{2}_{s, \infty} \Gamma}\,\rho_{\rm in}\,,
\end{equation}
where we have made use of the fact that initially the fluid is isentropic
and hence it can also be described by a polytropic equation of state
\citep{Cruz2012}.

\begin{figure*}
  \center
  \includegraphics[width=1.9\columnwidth]{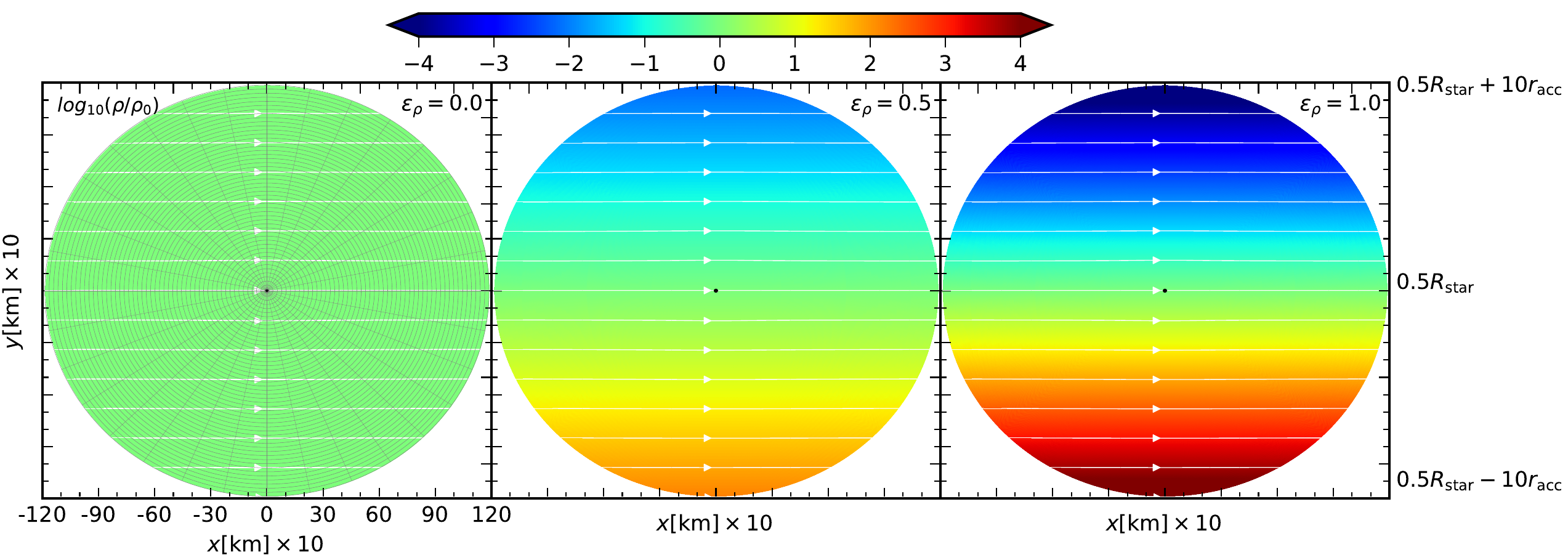}
  \caption{Initial distributions of the rest-mass density and the
    velocity field streamlines in the common envelope projected on the $x-y$
    plane for models $\texttt{RCE.0.0.5o3}$, $\texttt{RCE.0.5.5o3}$ and
    $\texttt{RCE.1.0.5o3}$ with adiabatic index $\Gamma=5/3$ and for the
    dimensionless density gradients $\epsilon_{\rho}=0.0, 0.5$, and  $1.0$,
    respectively. {In the left panel we overplot some of the grid lines
      ($1/40$ and $1/8$ in the radial and azimuthal direction,
      respectively) used in the numerical simulations. All axes have
      units of $\rm km$, with the exception of the vertical right
      vertical axis, which is expressed in terms of radius of the star
      and of the accretion radius $r_{\rm acc}$. Note that the black hole
      is located at the center of the numerical domain (shown with a
      black circle), which covers only a small fraction of the star
      radius $20\, r_{\rm acc} \sim 4.3 \times 10^{-6} R_{\rm star}$.}}
    \label{fig:Initdata}
\end{figure*}

The initial distributions of the rest-mass density of the common envelope
are plotted in Fig. \ref{fig:Initdata}, from right to left
$\epsilon_{\rho}=0.0, 0.5$ and $1.0$, which can be interpreted as
different stages of the common-envelope evolution. More specifically, as
the distance between the black hole and the center of the supergiant
decreases, the density gradients increase, thus going from
$\epsilon_{\rho} \sim 0.0$ to $\epsilon_{\rho} \sim 1.0$, where the
differences in the density at $y$-axis are $\Delta \rho=0$ to $\Delta
\rho=10^{8}$, respectively, while the streamlines (depicted with white
lines) show a uniform velocity in the $x$-direction.

\begin{table}
  \begin{center}
    \caption{Summary of the models considered, all of which refer to a
      Schwarzschild black hole, have a fixed sound speed of $c_{s,
        \infty}=0.1$, and an asymptotic Mach number ${\cal
        M}_{\infty}=2$, corresponding to an asymptotic velocity
      $v_{\infty}=0.2$; {this choice fixes the accretion radius $r_{\rm
          acc}$ (in mass units and in kilometers) to the value reported
        in the fourth column.} Hence, the parameters varied are the
      values of the adiabatic index $\Gamma$ and of the rest-mass density
      relative scale height $\epsilon_{\rho}$. {Finally, shown in the
        last column is the resolution in the radial direction $\Delta
        r$.}}
\begin{tabular}{lccccccc}
\hline\hline
 Model  & $\epsilon_{\rho}$ & ${\Gamma}$ & $r_{\rm acc}  [M \, ( {\rm km}) ]$&$\Delta r  [  M \, ( {\rm km}) ]$\\
\hline
\hline
$\texttt{RCE.0.0.5o3}$   &  $0.0$  &  $5/3$  & $20 \, (119)$ & $0.1 \, (0.6)$  \\
\hline
$\texttt{RCE.0.1.5o3}$   &  $0.1$  &  $5/3$  & $20 \, (119)$ & $0.1 \, (0.6)$\\
\hline
$\texttt{RCE.0.3.5o3}$   &  $0.3$  &  $5/3$  & $20 \, (119)$ & $0.1 \, (0.6)$\\
\hline
$\texttt{RCE.0.5.5o3}$   &  $0.5$  &  $5/3$  & $20 \, (119)$ & $0.1 \, (0.6)$\\
\hline
$\texttt{RCE.0.7.5o3}$   &  $0.7$  &  $5/3$  & $20 \, (119)$ & $0.1 \, (0.6)$\\
\hline
$\texttt{RCE.1.0.5o3}$   &  $1.0$  &  $5/3$   & $20 \, (119)$ & $0.1 \, (0.6)$\\
\hline
\hline
$\texttt{RCE.0.0.4o3}$   &  $0.0$  &  $4/3$   & $20 \, (119)$ & $0.1 \, (0.6)$\\
\hline
$\texttt{RCE.0.1.4o3}$   &  $0.1$  &  $4/3$   & $20 \, (119)$ & $0.1 \, (0.6)$\\
\hline
$\texttt{RCE.0.3.4o3}$   &  $0.3$  &  $4/3$   & $20 \, (119)$ & $0.1 \, (0.6)$\\
\hline
$\texttt{RCE.0.5.4o3}$   &  $0.5$  &  $4/3$   & $20 \, (119)$ & $0.1 \, (0.6)$\\
\hline
$\texttt{RCE.0.7.4o3}$   &  $0.7$  &  $4/3$  & $20 \, (119)$ & $0.1 \, (0.6)$\\
\hline
$\texttt{RCE.1.0.4o3}$   &  $1.0$  &  $4/3$   & $20 \, (119)$ & $0.1 \, (0.6)$\\
\hline\hline
\end{tabular}
\label{tab:params}
\end{center}
\end{table} 

Finally, when discussing the boundary conditions, we recall that the inner
boundary $r_{\rm exc}$ is always situated inside the event horizon, where
we apply an inward extrapolation of the conservative variables in the
state vector $\boldsymbol{U}$. On the other hand, on the external
boundary at $r_{\rm max}$ we impose a steady inflow of matter having the
same properties in terms of density, pressure, and velocity given by the
initial conditions, assuming an inflow in the upstream part of the
computational domain, \ie for $\pi/2 < \phi < 3\pi/2$, and an outflow in
the downstream part, \ie for $-\pi/2 < \phi < \pi/2$. Finally, we impose
simple periodic boundary conditions in the $\phi$-direction.

\section{Results}
\label{sec:Res}

\subsection{Morphology of the accreting plasma}
\label{ss:Morpho}

\begin{figure*}
  \center
  \includegraphics[width=1.75\columnwidth]{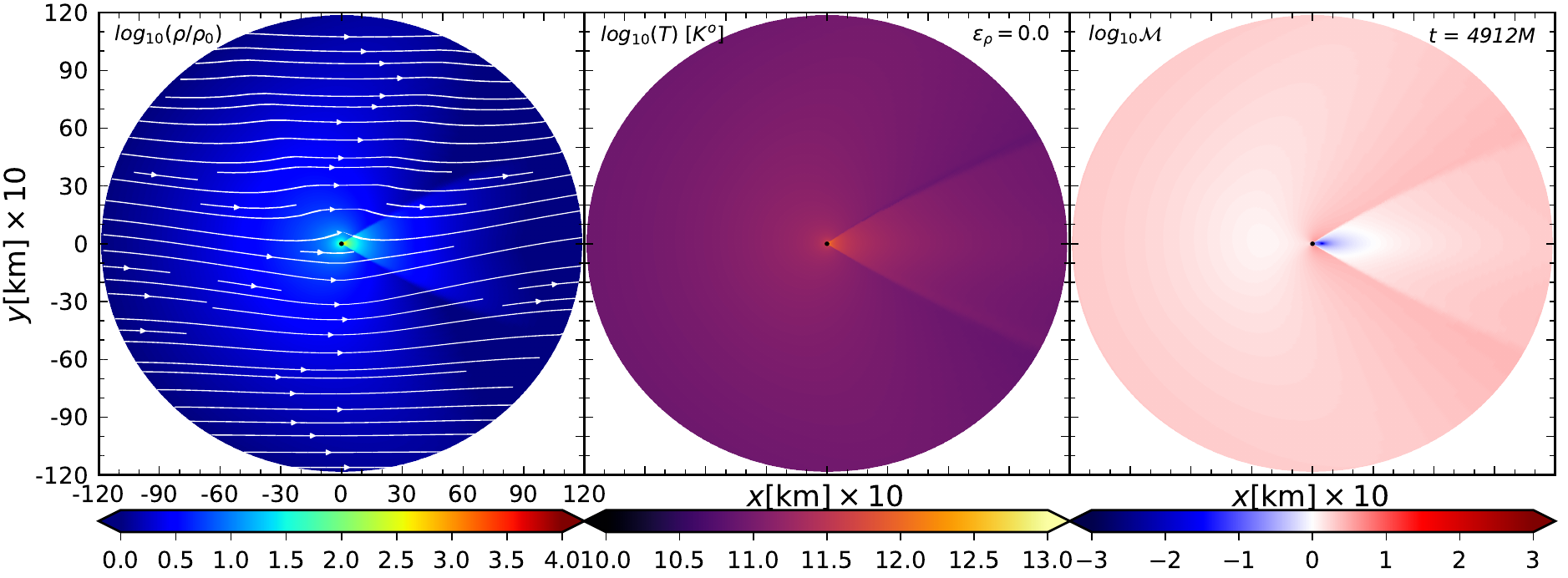}\\
  \includegraphics[width=1.75\columnwidth]{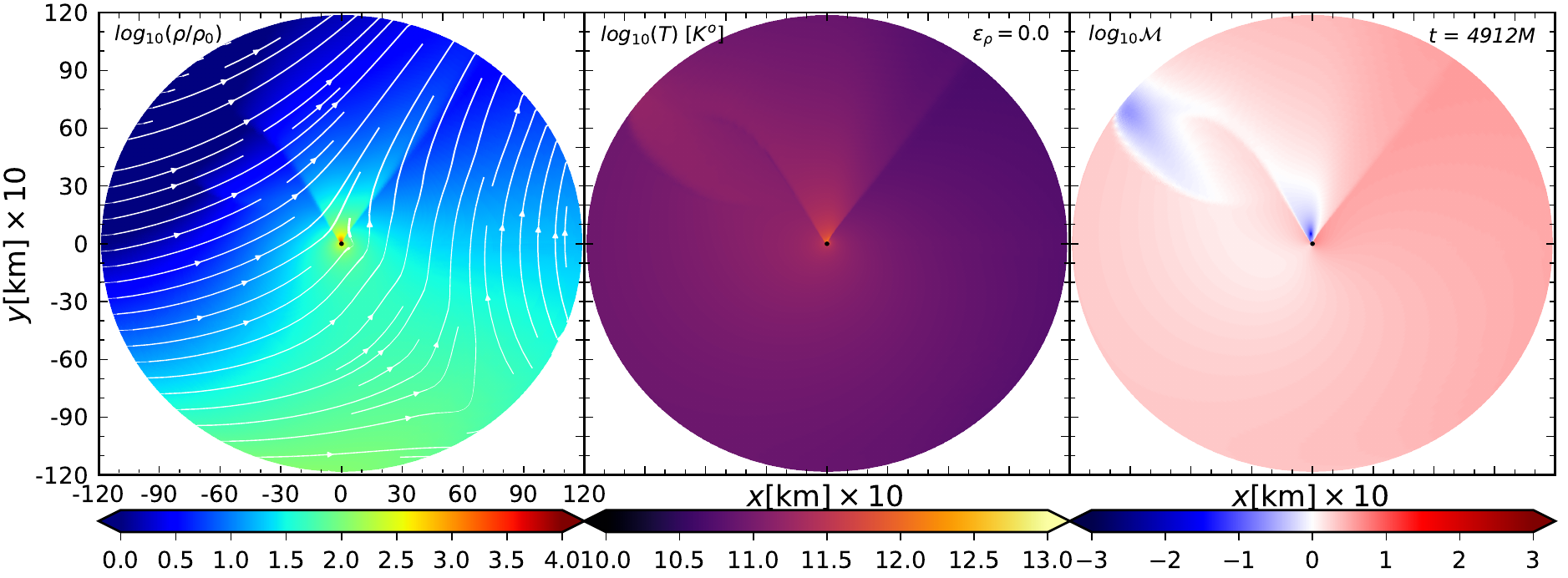}\\
  \includegraphics[width=1.75\columnwidth]{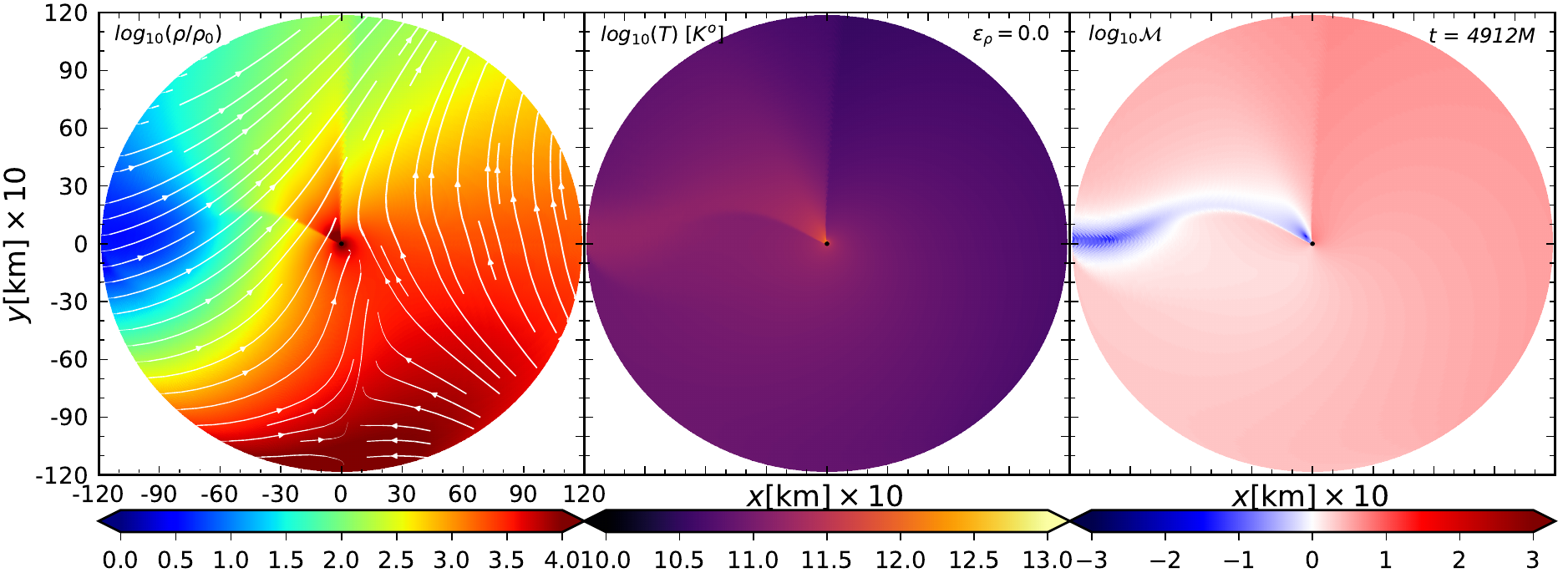}
  \caption{Snapshots of the rest-mass density, temperature, and Mach
    number at the end of our numerical simulations $t\sim 5000\, M \sim
    0.1\,{\rm s}$ for models $\texttt{RCE.0.0.4o3}$,
    $\texttt{RCE.0.5.4o3}$ and $\texttt{RCE.1.0.4o3}$. From top to
    bottom, we show the morphology of the accreting matter envelope for
    dimensionless density gradients of $\epsilon_{\rho=}0, 0.5$, and  $1.0$. In
    all panels, the adiabatic index is set to $\Gamma=4/3$.}
\label{fig:morphology1}
\end{figure*}

\begin{figure*}
  \center
  \includegraphics[width=1.75\columnwidth]{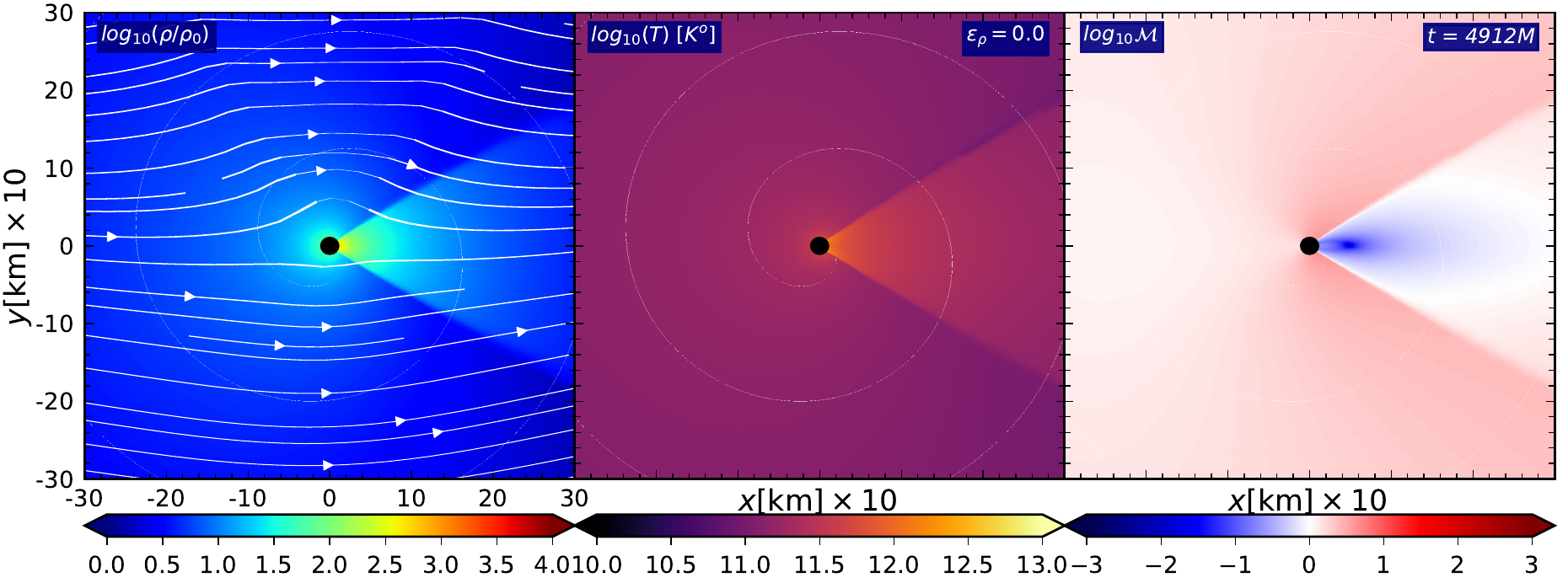}\\
  \includegraphics[width=1.75\columnwidth]{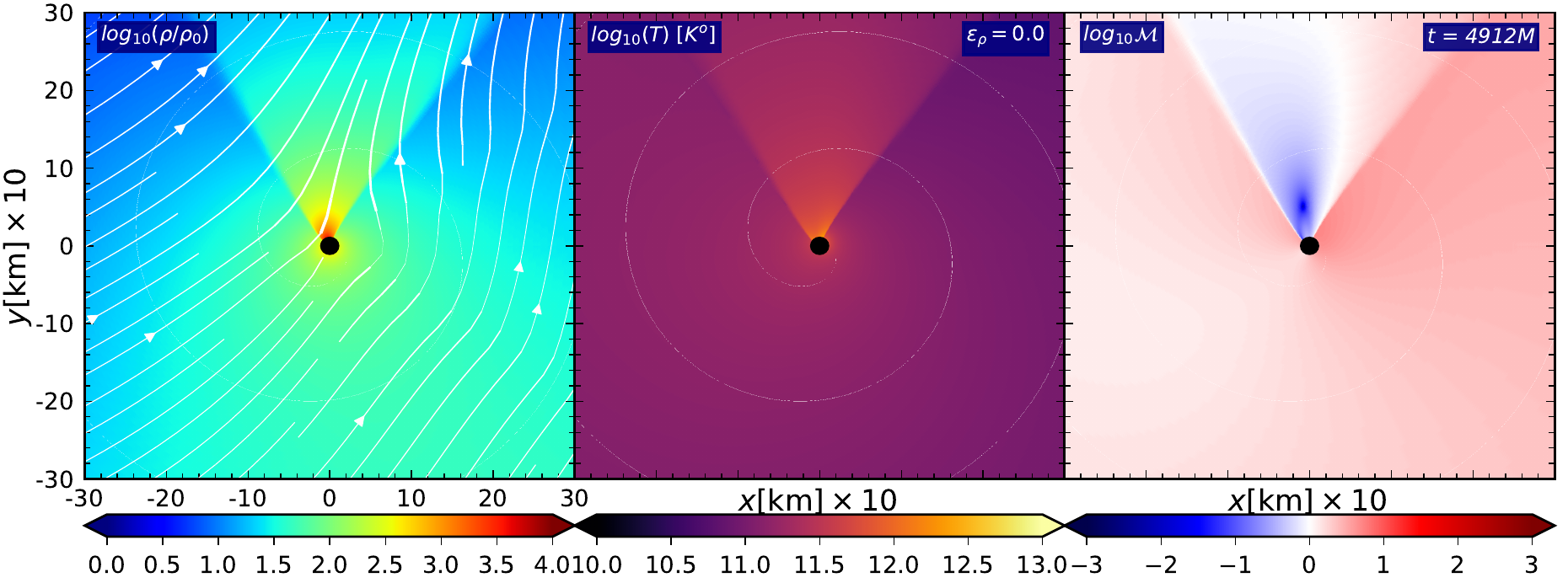}\\
  \includegraphics[width=1.75\columnwidth]{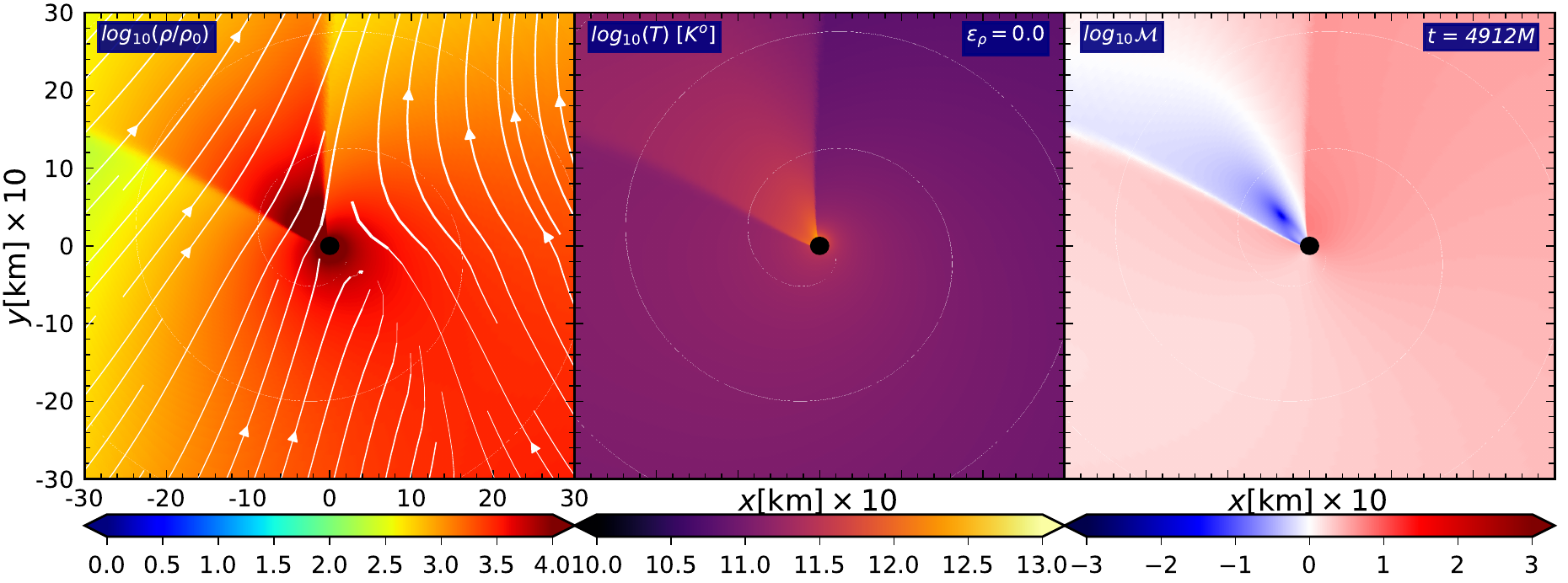}
  \caption{Close-up view of the various panels of
    Fig. \ref{fig:morphology1} to highlight the dynamics near the
    accreting black hole.}
\label{fig:morphology1_z}
\end{figure*}

The morphology of the evolved common envelope as sketched in the cartoon
in Fig. \ref{fig:tzo} -- black hole moving perpendicular to the direction
of the gradient -- is shown in Fig.  \ref{fig:morphology1}, where we
report, for the $\Gamma=4/3$ case, the different panels: the rest-mass
density (left panel), the temperature (left panel), and the Mach number
$\mathcal{M}:=Wv/W_sc_s$, where $W_s$ is the Lorentz factor of the sound
speed (right panel). The snapshots refer to $t \sim 5000 M\, \sim 0.1
\,{\rm s}$, which is obviously much shorter than the typical evolution
timescale of a common-envelope scenario. On the other hand, this time
corresponds to $\sim 50$ crossing times\footnote{We define the crossing
  time as $t_{\rm cross}:= r_{\rm acc}/v_{\infty}$, so that the spatial
  computational domain is covered in $\sim 10 t_{\rm cross}$.}, 
  when the gas has reached a steady state and serves therefore as a
useful reference for the stationary dynamics of the system. More
specifically, the different rows of Fig. \ref{fig:morphology1} refer to
different values of the dimensionless density gradient, \ie
$\epsilon_{\rho}=0,\ 0.5$ and $1.0$ from top to bottom, where the case
with $\epsilon_{\rho}=0$ corresponds to a constant rest-mass density
profile. A magnified view of the same panels is offered in
Fig. \ref{fig:morphology1_z} and helps us appreciate the dynamics of the
flow in the vicinity of the black hole.

\begin{figure*}
  \center
  \includegraphics[width=1.75\columnwidth]{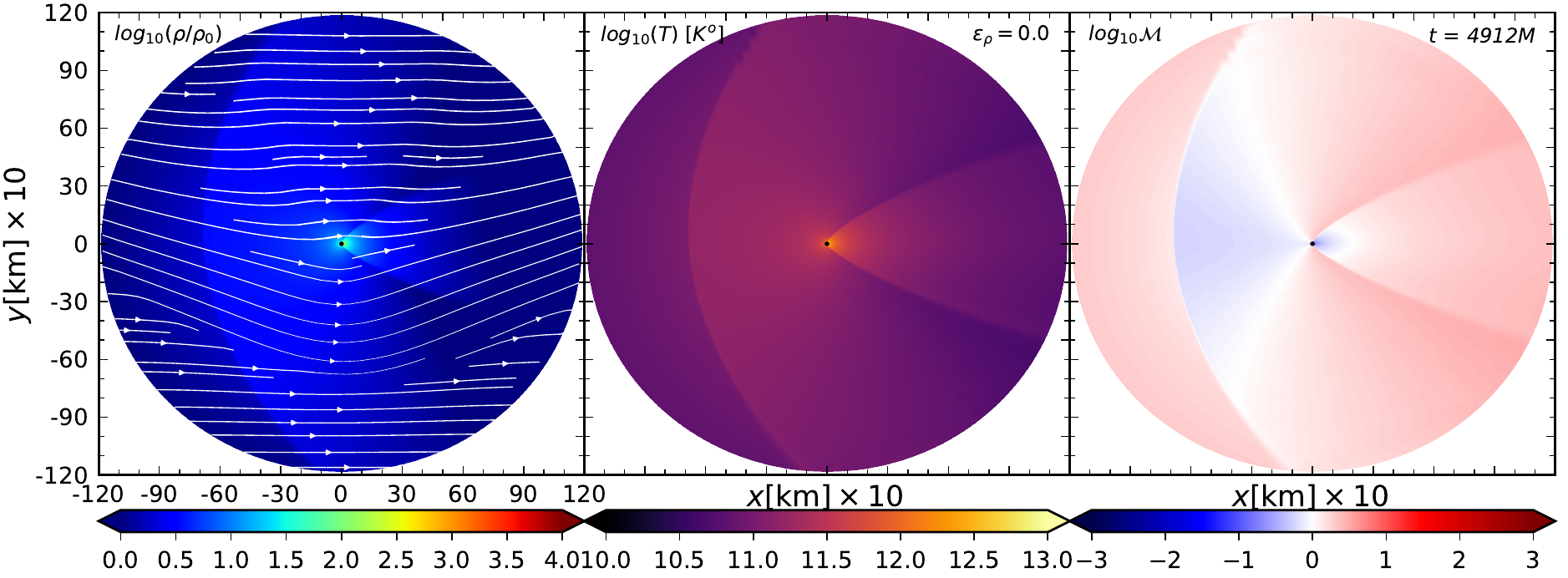}\\
  \includegraphics[width=1.75\columnwidth]{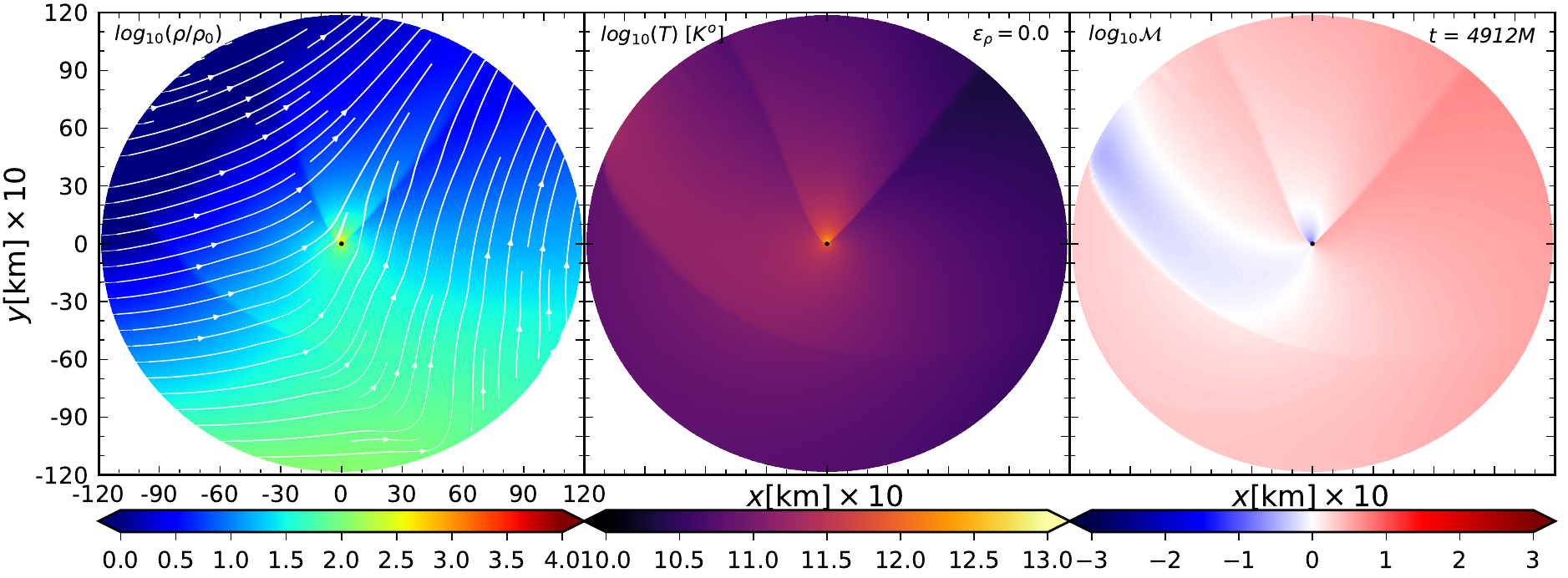}\\
  \includegraphics[width=1.75\columnwidth]{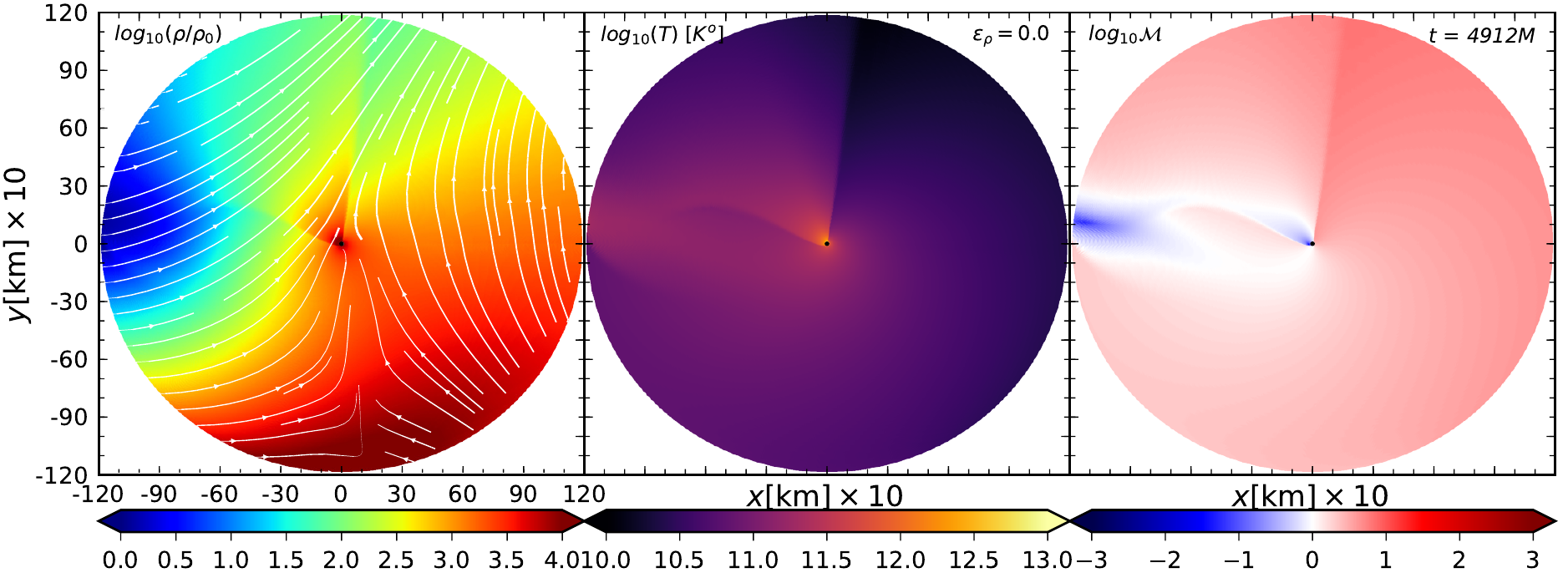}
  \caption{Same as in Fig. \ref{fig:morphology1}, but for models
    $\texttt{RCE0.0.5o3}$, $\texttt{RCE.0.5.5o3}$, and
    $\texttt{RCE.1.0.5o3}$. In all panels, the adiabatic index is set to
    $\Gamma=5/3$.}
  \label{fig:morphology2}
\end{figure*}

The temperature in the central column of {Figs.}  \ref{fig:morphology1}
to \ref{fig:morphology2} was computed using the definition from
\citep{Zanotti2010},
\begin{eqnarray}
T&=& 1.088 \times 10^{13} \left( \frac{p}{\rho} \right) K\,,
\label{eq:temp}
\end{eqnarray}
where the numerical factor comes from the transformation from geometric
units to Kelvin.

We start by considering the behaviour of the accretion of a fluid with
$\Gamma=4/3$ as reported in Figs. \ref{fig:morphology1} and
\ref{fig:morphology1_z}. The dynamics observed in the case of the motion
across a constant-density cloud reproduces well the behavior observed in
simulations of Bondi--Hoyle accretion, where the shock cone in the wake of
the black hole is stable and symmetric with respect to the to direction
of motion, \ie the $x$-axis in this case 
{\citep[see, e.g.,][]{Font1999b, Donmez2010, Cruz2012, Lora2015219} }. 
This is highlighted in {Figs. \ref{fig:morphology1} and \ref{fig:morphology1_z}
  by the streamlines of the flow (white lines)} and by the moderate
compression in the matter density around the black hole, which shows an
increase of less than two orders of magnitude in the rest-mass density.

On the other hand, the dynamics produced in models $\texttt{RCE.0.5*}$
and $\texttt{RCE.1.0*}$ shows that, once formed, the shock cone is
stationary but tends to be directed along the direction of the gradient
in the rest-mass density, and hence pointing away from the region of high
density in the cloud, namely, the center of the red supergiant. Note,
however, that as the dimensionless accretion radius is increased, the
density profile along the direction of motion of the black hole develops
an asymmetry, with a low-density region ahead of the black hole and a
corresponding high-density region behind. As a result, the shock cone --
which is aligned along the direction of motion (\ie the horizontal
direction in our setup) for a uniform medium -- is now dragged by the
density gradient and tends to align with it for $\epsilon_{\rho}=0.5$.

This is shown in the left panels of Figs. \ref{fig:morphology1} and
\ref{fig:morphology1_z}, both of which refer to the case with
$\Gamma=4/3$ and show different scales of the computational domain. In
these figures, from top to bottom, the gradual increase in the density
gradient -- which can be taken to mimic the motion of the black hole as
it gets closer to the core of the red supergiant -- leads to an
increasingly large angle with respect to the direction of motion (see
the diagram in Fig. \ref{fig:tzo}), which can be larger than $\pi/2$ for the
case of the largest value of the initial gradient $\epsilon_{\rho}=1.0$
(bottom row). Interestingly, in the latter case, most of the motion
downstream of the black hole is actually away from the core of the
supergiant.

The motion of the matter is best followed through the streamlines of
fluid elements and shown in the left column of
Figs. \ref{fig:morphology1} and \ref{fig:morphology1_z}. When contrasting
the top and bottom rows, it is possible to appreciate that while the fluid
elements always move in the positive $x$-direction in the case of a
medium that is uniform or with a small density gradient (top and middle
row), they can also move in the negative $x$-direction in the case of a
large density gradient (bottom row). However, {in contrast with what
 was  reported in the Newtonian simulations} of \citet{MacLeod2017}, no
significant vorticity is found in these regimes; while a vortical motion
may well be produced at even larger density gradients, the stronger
general relativistic gravitational fields prevent their formation here.

The development of the shock cone heats up the fluid, leading to a
temperature increase that is about one order of magnitude larger than in the
rest of the envelope, with a fluid that is overall two orders of
magnitude hotter than in the initial state (see central columns in
Figs. \ref{fig:morphology1} and \ref{fig:morphology1_z}). As is well known
in the phenomenology of Bondi--Hoyle--Lyttleton accretion across a uniform
medium, a stagnation point develops in the shock cone, with some of the
material into the shock cone accreting subsonically onto a black hole (\ie
${\cal M} \sim 10^{-1}$). These regions are shown in blue in the right
columns of Figs. \ref{fig:morphology1} and \ref{fig:morphology1_z}, which
refer to the Mach number. Interestingly, the dragging of the shock cone
also leads to the formation of a stagnating subsonic flow in the presence
of a density gradient, which becomes increasingly severe as the gradient
is increased (middle and bottom rows). In this case, large variations in
the Mach number are {produced}, with variations of almost two orders of
magnitude.

When changing the adiabatic index, that is, when considering the same
setups but for an ideal-fluid equation of state with $\Gamma=5/3$, and hence
considering a different compressibility of the gas in the
common envelope, no major qualitative differences appear in the overall
properties of the flow, although quantitative differences do emerge and
will be measured in the following sections. Besides slightly larger
temperatures (of about a factor of two), the most significant difference
that emerges in simulations with $\Gamma=5/3$ is the clear appearance of
a bow shock in the upstream region of the flow and that follows the
rotations of the shock cone as the density gradient is increased (see
Fig. \ref{fig:morphology2}). Before concluding this section on the
overall dynamics of the flow, we should note that in our simulations the
bow shock appears only for $\Gamma=5/3$, unlike in the Newtonian case
\citep{MacLeod2015, MacLeod2017}, where it is present also for
$\Gamma=4/3$.

\subsection{Mass accretion and drag rates}
\label{sec:ma_dr}

Besides a general understanding of the dynamics of the plasma as it
interacts with the black hole -- which has an interest of its own -- our
simulations are meant to also measure the typical rates of accretion of
mass and momentum, since the latter have important implications on the
quantitative description of the common-envelope evolution. In the case of
spherical accretion onto a black hole and an ideal-fluid equation of
state, analytic expressions can be obtained for the accretion rates of
mass and momentum \citep{Petrich89, Rezzolla_book:2013}
\begin{eqnarray}
  \label{eq:canmdot}
  \dot{M}_{\rm ref}&=& 4 \pi \lambda \rho_{\infty} M^{1/2}r^{3/2}_{\rm
    acc}\,,\\
  \label{eq:canpdot}
  \dot{P}_{\rm ref}&=& \dot{M}_{\rm ref}\frac{v_{\infty}}{\sqrt{1-v^{2}_{\infty}}}\,,
\end{eqnarray}
where
\begin{eqnarray}
  \lambda &:=& \left(\frac{1}{2}\right)^{(\Gamma +1)/(2(\Gamma -1))}
  \left(\frac{5-3\Gamma}{4}\right)^{-(5-3\Gamma)/(2(\Gamma-1))}\nonumber\\
  &\simeq& \left\{
  \begin{array}{lr}
    0.71 & \qquad (\Gamma=4/3) \\
    0.25 & \qquad (\Gamma=5/3)
  \end{array}
  \right.\,.
  \label{eq:canlambda}
\end{eqnarray}

Equations \eqref{eq:canmdot} and \eqref{eq:canpdot} will be used
hereafter as a reference values to express the numerically computed
mass accretion and drag rates (also referred to as ``momentum accretion
rates''), which we compute as \citep{Petrich89}
\begin{eqnarray}
\label{eq:mdot}
\Dot{M} &=& \int_{0}^{2\pi} \alpha\sqrt{\gamma}D(v^{r} - \beta^{r}/\alpha) d\phi\,,\\
\Dot{P}^{i}&=&-\int_{\partial V} \alpha \sqrt{\gamma} T^{i j} d\Sigma_{j}
+ \int_{V} \alpha \sqrt{\gamma} S^{i} dV\,.
\label{eq:pdot}%
\end{eqnarray}
where $i=r, \phi$, and the inward fluxes are measured at the event
horizon.

\begin{figure}
  \center
  \includegraphics[width=0.90\columnwidth]{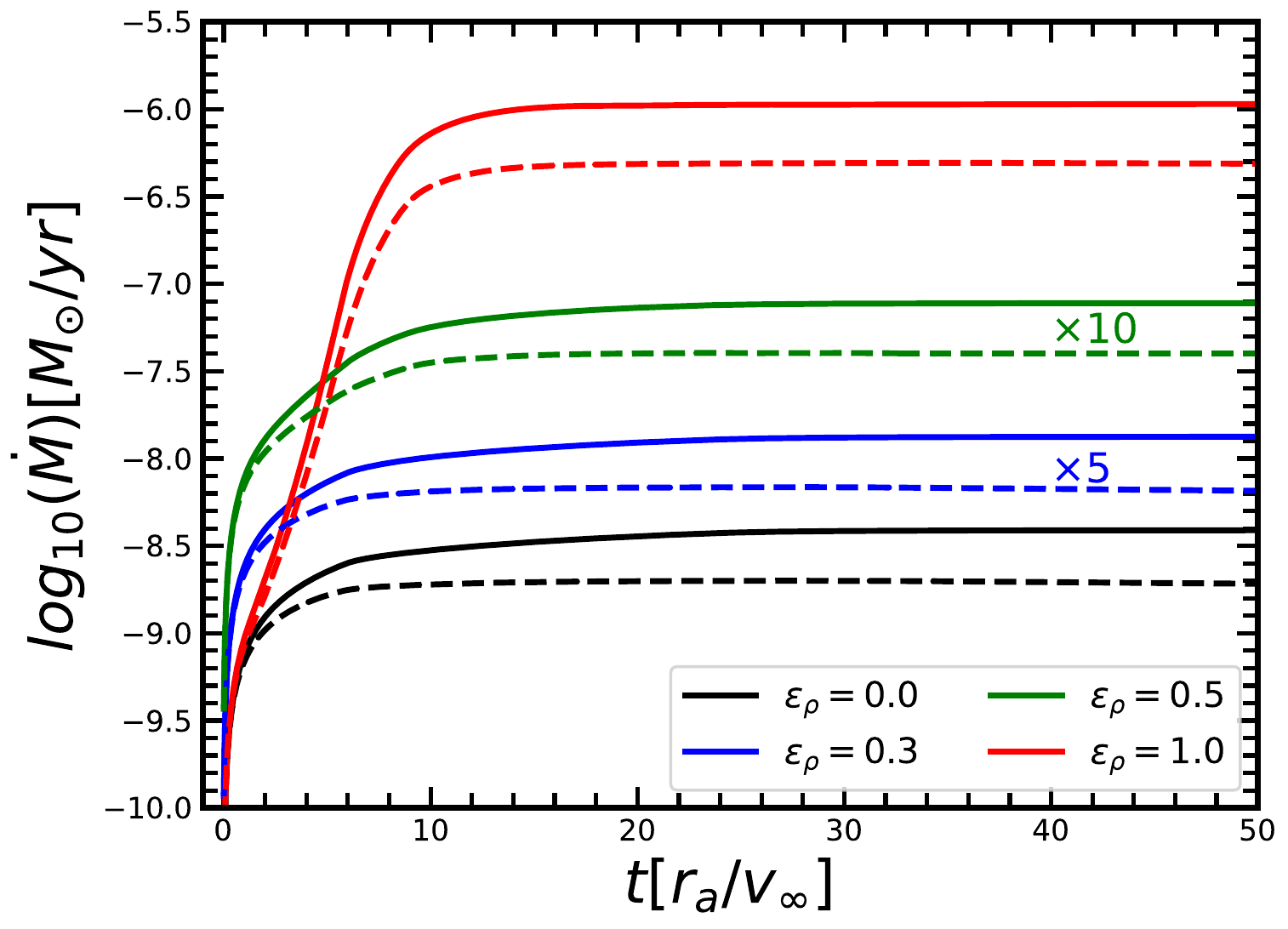}
  \caption{Evolution of the mass accretion rates for the adiabatic
    indices $\Gamma=4/3$ (solid lines) and $\Gamma=5/3$ (dotted
    lines). In all models, the accretion process reaches a stationary
    state after about 10 crossing times. Note that the
    low-$\epsilon_{\rho}$ curves are rescaled to appear on the same
    plot.}
    \label{fig:ratesvstime}
\end{figure}

\begin{figure}
  \center
  \includegraphics[width=0.99\columnwidth]{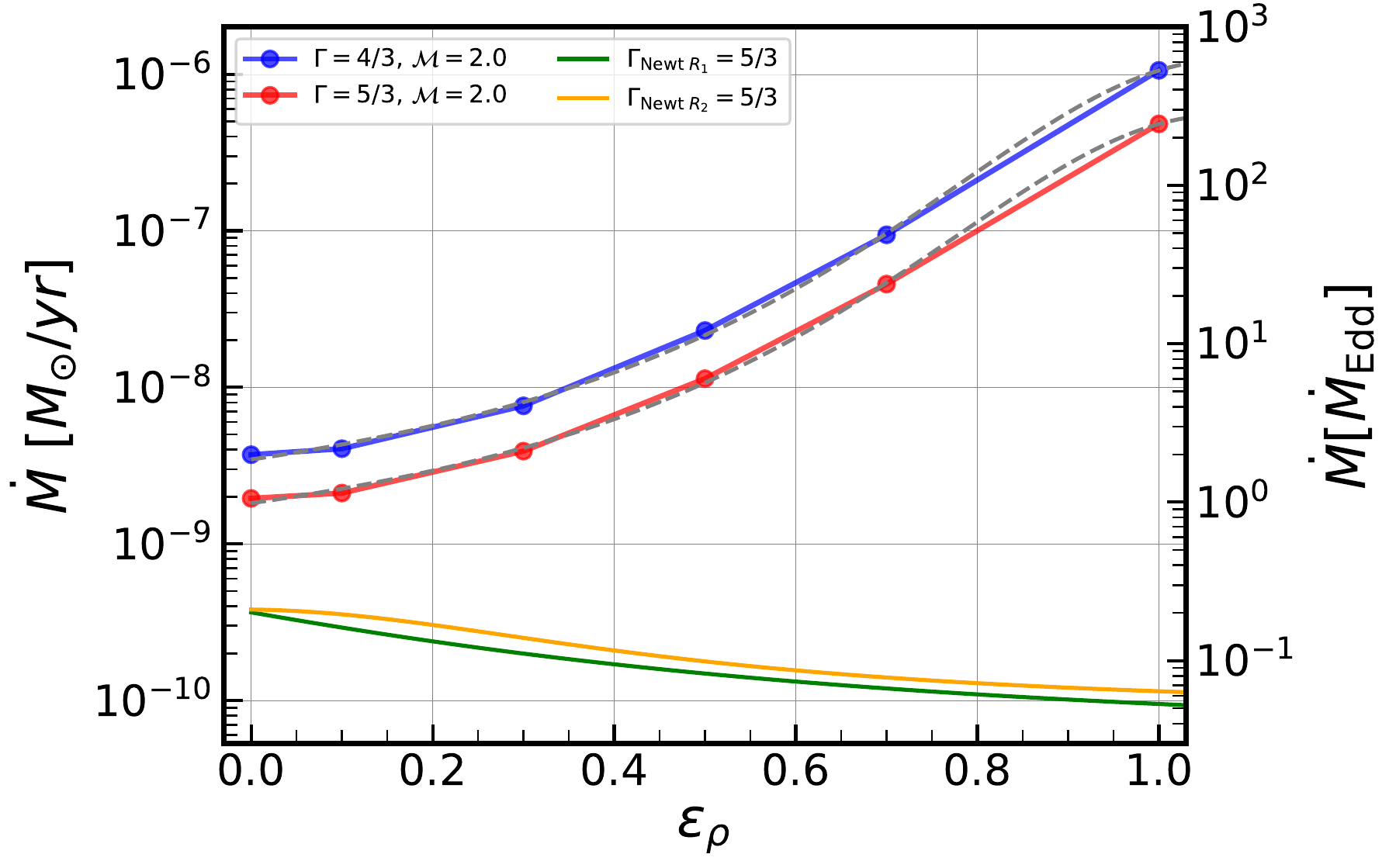}
  \caption{Mass accretion rates -- expressed either in solar masses per
    year or in Eddington rates -- computed at the event horizon and shown
    as a function of the dimensionless density gradient
    $\epsilon_{\rho}$. Shown with blue and red lines are the results for
    a fluid with $\Gamma=4/3$ and $\Gamma=5/3$, respectively. Also shown
    are the corresponding rates as reported in Newtonian simulations and
    when computed at two different radii, \ie $R_{1}=0.01 r_{\rm acc}$
    and $R_{1}=0.05 r_{\rm acc}$ \citep{MacLeod2015}. Note that while
    the general relativistic rates grow almost exponentially with
    $\epsilon_{\rho}$, the Newtonian ones decrease.}
    \label{fig:rates}
\end{figure}

In Fig. \ref{fig:ratesvstime} we report the mass accretion rates as a
function of time for the various models considered. Note that all models
reach a stationary accretion state after about 20 crossing times, and
we use such asymptotic values to construct the data shown in
Fig. \ref{fig:rates}, which provides a summary of the mass accretion for
all of the cases considered in terms of the dimensionless density
gradient $\epsilon_{\rho}$ and equation of state; also reported for
comparison are the values presented by \cite{MacLeod2015} for their
simulations in Newtonian gravity. Note that for configurations with a
zero density gradient ($\epsilon_{\rho}=0$), we reproduce the expected
super-Eddington accretion rate ($\dot{M} \sim 10^{-8}$) for
Bondi--Hoyle--Lyttleton accretion obtained also in previous simulations
\citep{Zanotti2010,Zanotti2011,Lora2015219}; furthermore, these values
have been cross-checked also with a different general relativistic MHD
code (see discussion in Appendix \ref{sec:appB}). Figure \ref{fig:rates2}
reports a very similar behavior also for the rates of accretion of
radial momentum (left panel) and of angular momentum (right panel).

Figure \ref{fig:rates} also shows that, for non-negligible density
gradients, \ie $\epsilon_{\rho} \gtrsim 0.3$, both the mass accretion
rate and the drag rate grow exponentially with the density gradient and
can therefore be well approximated with functions of the type
\begin{eqnarray}
  \log\left(\frac{\dot{M}}{\dot{M}_{\rm ref}}\right)&=& \mu_{1} +
  {\mu_{2}}/({1+\mu_{3}\epsilon_{\rho} + \mu_{4}\epsilon^{2}_{\rho}})\,,
  \label{eq:fitmrate}\\
  \log \left(\frac{\dot{P}^{\phi}}{\dot{P}_{\rm ref}}\right) &=&
  \wp_{1} +\wp_{2}\epsilon_{\rho} + \wp_{3}\epsilon^{2}_{\rho}\,,
\label{eq:fitmom}
\end{eqnarray}
where the values of the fitting coefficients $\mu_{i}$ and $\wp_{i}$ are
reported in the Table \ref{tab:massscoef} for the two adiabatic indices
$\Gamma=5/3$ and $\Gamma=4/3$. Using expressions \eqref{eq:fitmrate} and
\eqref{eq:fitmom} it is possible to obtain mass accretion rates also when
considering different initial conditions as long as they are not very
different from those considered here (see also the discussion in
Appendix \ref{sec:iotid})\footnote{Using the scaling relations
    \eqref{eq:fitmrate} and \eqref{eq:fitmom} in very different regimes,
    \eg for initial velocities of the order of $10^{2}-10^{3}\,{\rm
      km/s}$ is of course possible, but this basically amounts to an
    extrapolation. As discussed in Appendix \ref{sec:iotid}, we have
    verified that the scaling relations work very well when reducing
    $v_{\infty}$ by a factor of two, reproducing accurately the results
    of \citet{Zanotti2011}; we expect this to be true also for much
    smaller values. A more systematic analysis of the dependence of the
    accretion rate on $\mathcal{M}_{\infty}$ will be part of our future
    work.}. For example, given $\left({\dot{M}}/{\dot{M}_{\rm
    ref}}\right)_{\rm old}$ as expressed by Equation \eqref{eq:fitmrate}, if one
wishes to calculate a new mass accretion rate $(\dot{M})_{\rm new}$ using
a different initial velocity or rest-mass density, it is sufficient to
take into account the changes introduced by the new accretion radius in
the reference mass accretion rate $(\dot{M}_{\rm ref})_{\rm new}$ so that
$(\dot{M})_{\rm new} = (\dot{M}_{\rm ref})_{\rm new}
\left({\dot{M}}/{\dot{M}_{\rm ref}}\right)_{\rm old}$.

\begin{figure*}
  \center
  \includegraphics[width=0.75\columnwidth]{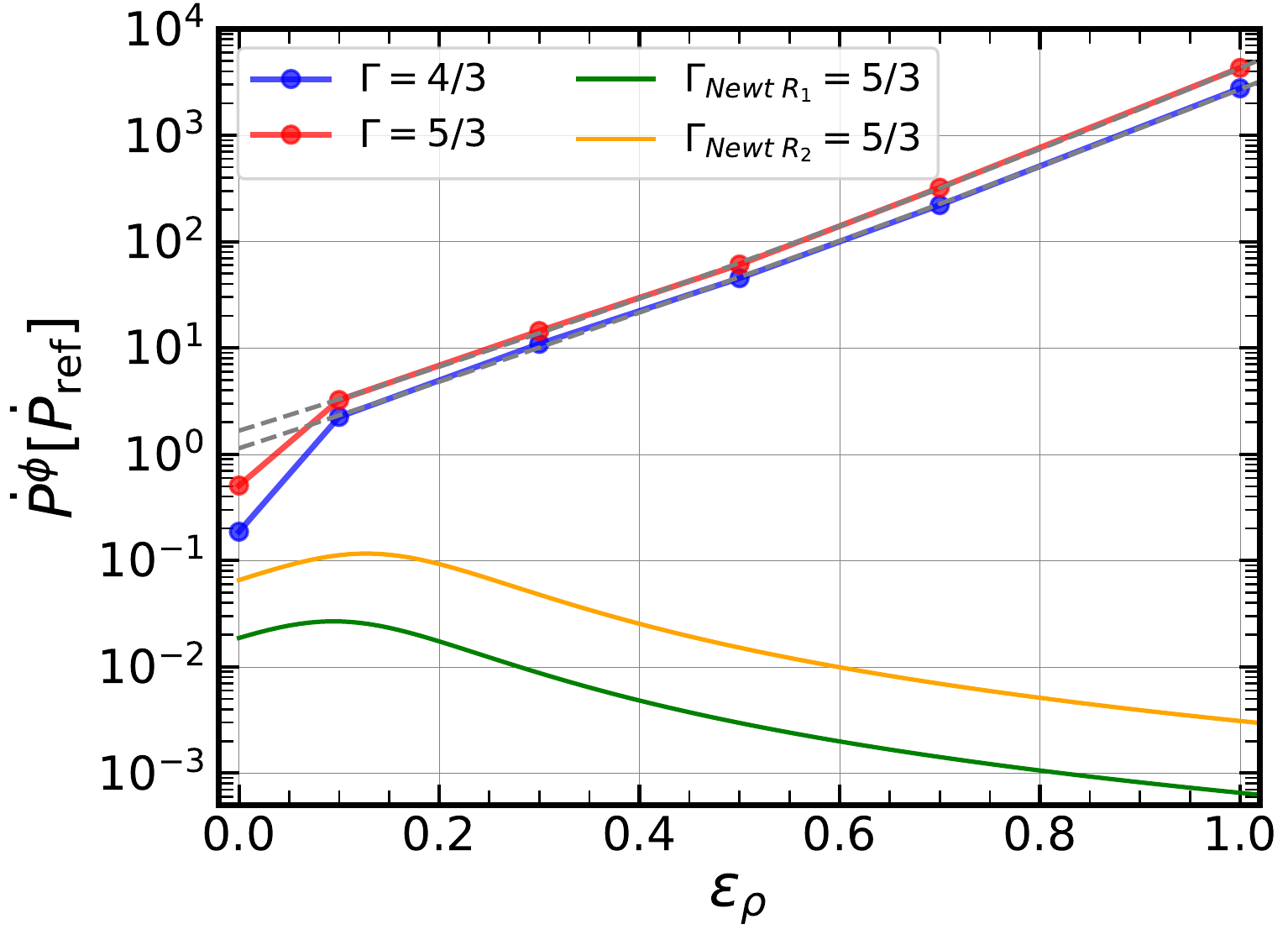}
  \hskip 2.0cm
  \includegraphics[width=0.75\columnwidth]{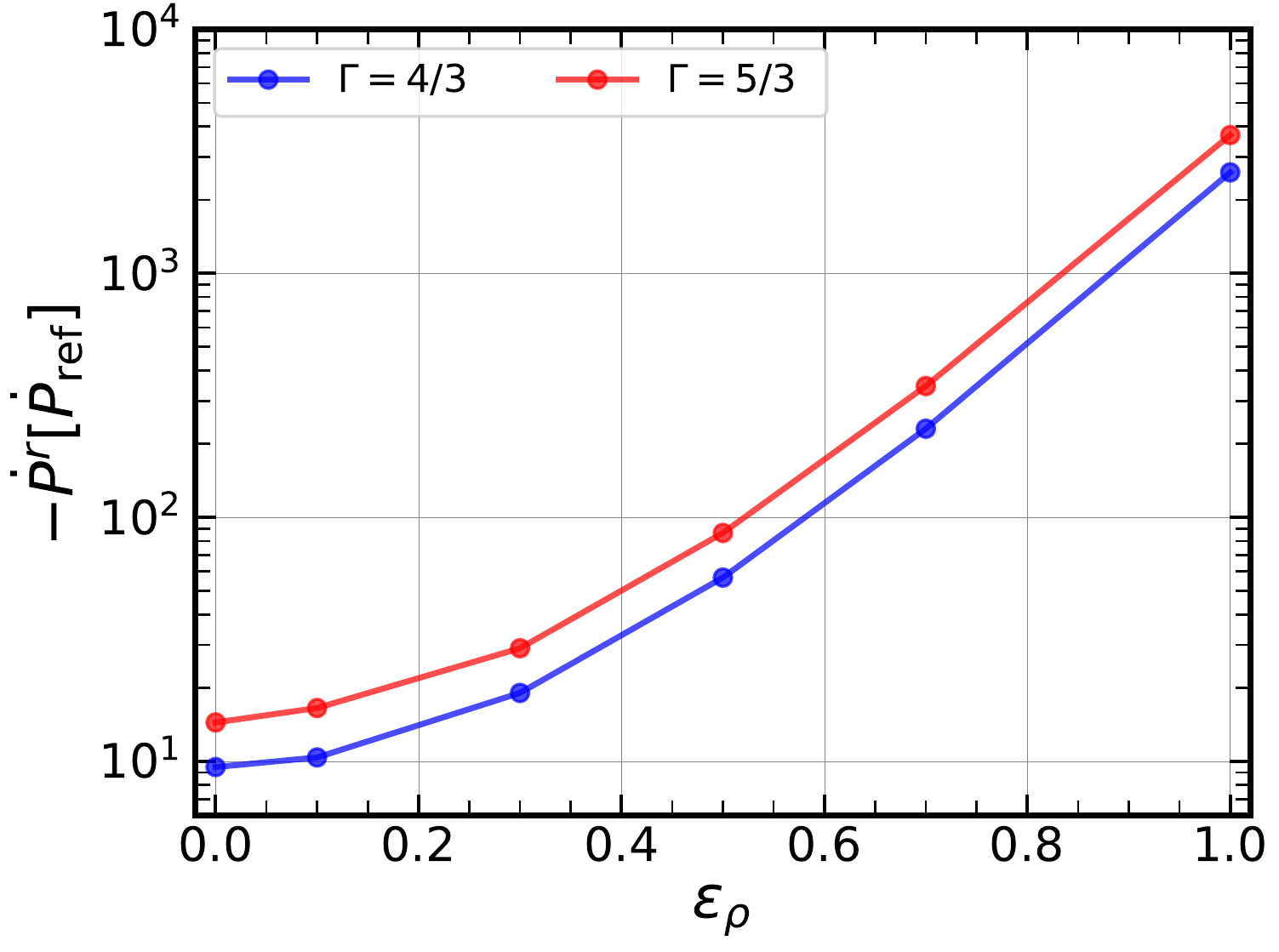}
  \caption{Same as in Fig. \ref{fig:rates} but for the accretion of
    angular momentum (left) and of radial momentum (right). Also in this
    case, note that while the general relativistic rates grow almost
    exponentially with $\epsilon_{\rho}$, the Newtonian ones decrease.}
    \label{fig:rates2}
\end{figure*}

\begin{figure}
  \center
  \includegraphics[width=0.75\columnwidth]{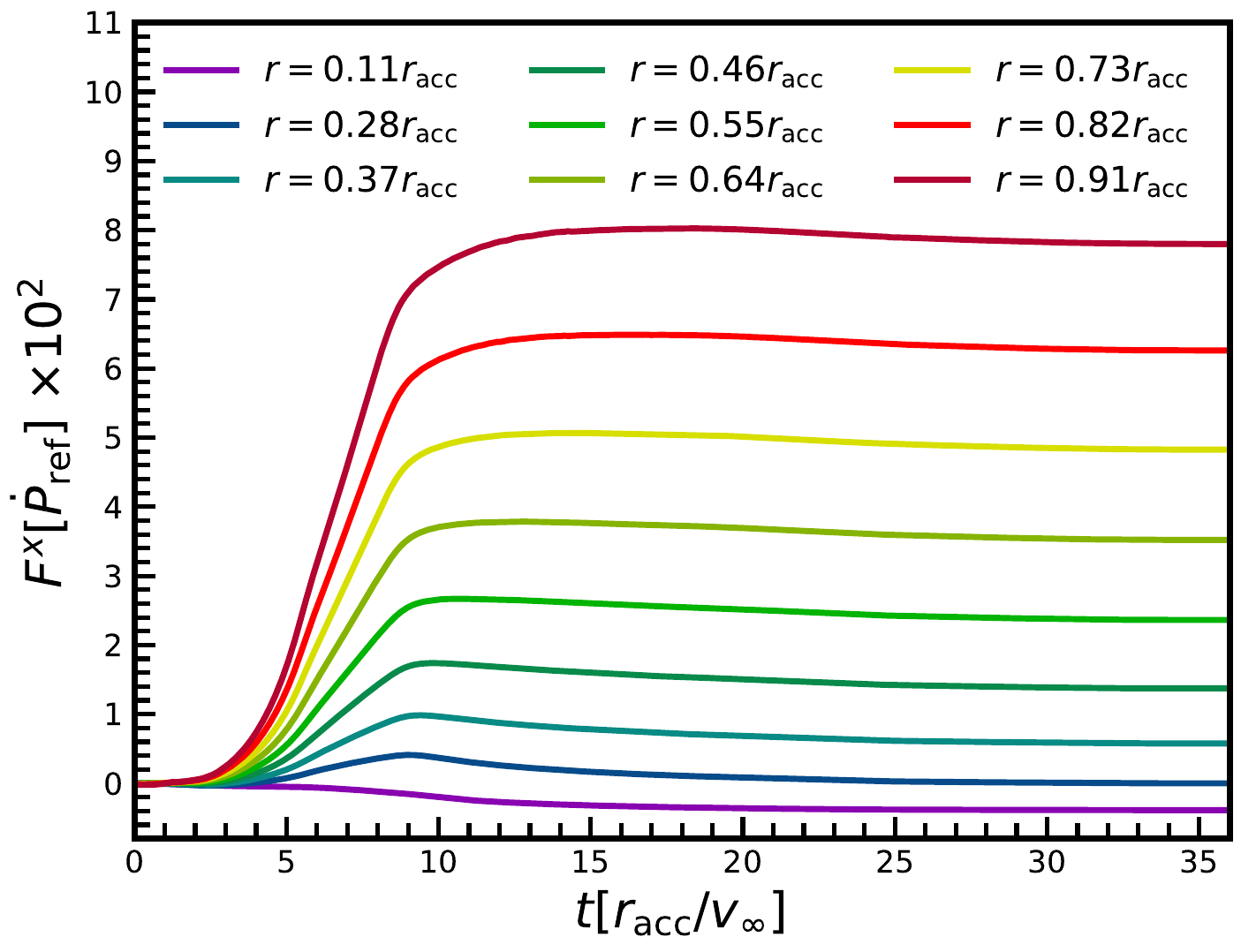}
  \includegraphics[width=0.75\columnwidth]{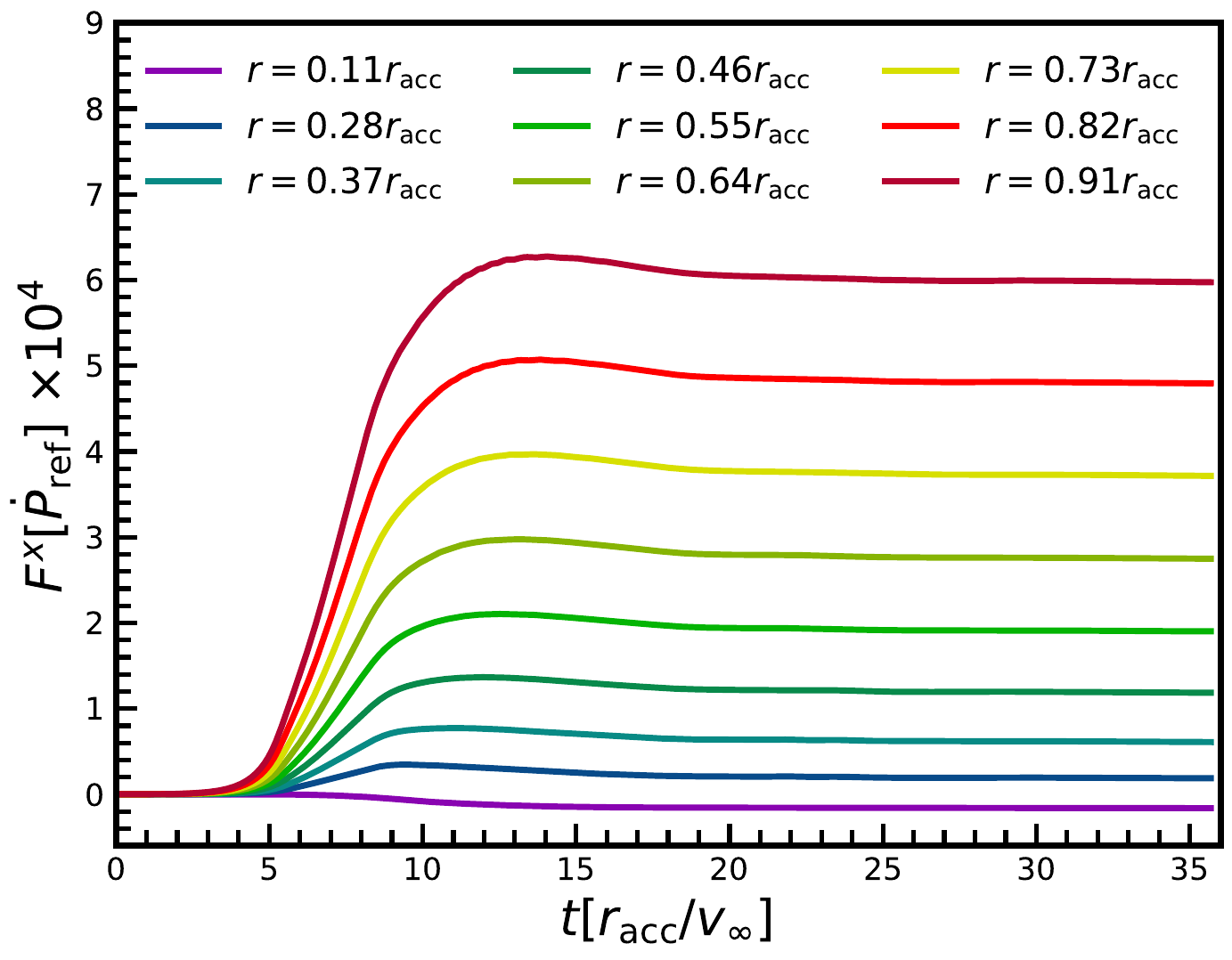}
  \caption{Total drag force along the $x$-direction (see Equation
   \eqref{eq:totalforce}) produced by the combination of the accretion
    of linear momentum and that of the dynamical friction. The different lines
    refer to the different spherical surfaces where the fluxes are
    computed (expressed in terms of the accretion radius), while the top
    and bottom panels refer to dimensionless density gradients of
    $\epsilon_{\rho}=0.5$ and $\epsilon_{\rho}=1.0$, respectively.}
    \label{fig:dragforcedetectors}
\end{figure}

When comparing the results of the mass accretion rates in the top panel
of Fig. \ref{fig:rates} and \ref{fig:rates2} with the corresponding
nonrelativistic values in Newtonian gravity \citep{MacLeod2015}, the
most striking difference is in the dependence from the density
gradient. While, in fact, the two accretion rates are comparable for
small gradients, the relativistic ones grow with $\epsilon_{\rho}$, the
opposite is true in the non-relativistic case. It is presently difficult
to ascertain the origin of this difference, but it may ultimately be
associated with the stronger gravitational fields that the fluid
experiences in our simulations. Indeed, a mass accretion rate that
increases with the density contrast is rather natural to expect since the
local rest-mass density near the black hole will be intrinsically
large. On the other hand, this behavior is not found in a Newtonian
regime because of the appearance there of disk-like structures with 
nonzero angular momentum that suppress the infall of the matter toward the
black hole. As mentioned above, no such vortical motions emerge in our
simulations.

Very similar considerations apply also when comparing {the relativistic
  and Newtonian drag rates} (bottom panel of Fig. \ref{fig:rates}). Also
in this case, in fact, the relativistic drag rates -- which are
ultimately responsible for the loss of orbital angular momentum in the
binary system and thus to the decrease in the orbital period
\citep{Taam2010} -- increase with the density gradient (as one would
naturally expect), while they decrease in the Newtonian case. Once again,
this different behavior may be due to the different role played by the
gravitational forces. On the other hand, it may also be due to the
different boundary conditions, which in our case are imposed inside the
event horizon by using the ingoing Eddington--Finkelstein coordinates and
are those of a purely infalling gas. A more involved description is
instead adopted in the Newtonian calculations of \cite{MacLeod2015},
where spherical absorbing boundary conditions are applied to a ``sink''
surrounding the central point mass, whose gravitational potential of the
point mass is smoothed within this sink.

\begin{table}
  \begin{center}
    \caption{Fitting coefficients for the mass accretion rate and the
      drag rate given by Eqs. (\ref{eq:fitmrate}) and (\ref{eq:fitmom}),
      respectively. }
\begin{tabular}{lccccccc}
\hline\hline
$\Gamma$   & $\mu_{1}$   &  $\mu_{2}$    &    $\mu_{3}$  &  $\mu_{4}$  &  $\wp_{1}$    &     $\wp_{2}$    &    $\wp_{3}$ \\
$4/3$      & $-0.29$  &  $1.39$  &  $-1.52$ &$0.70$  & $0.13$  &  $7.06$  &  $0.72$    \\
\hline                                                                                
$5/3$      & $0.18$  &  $1.24$   & $-1.54$  &$0.72$  & $0.51$  &  $6.69$   & $1.16$   \\
\hline
\hline
\end{tabular}
\label{tab:massscoef}
\end{center}
\end{table} 

\subsection{Drag forces, drag angle, and emitted radiation}

Drag forces experimented by the black hole as it moves in the envelope of the
red supergiant star are responsible for its deceleration. Estimating
these forces in a fully general relativistic context is of course
possible, but it is also less transparent than in a simpler Newtonian
description. In view of this, and in order to make the comparison with
previous Newtonian estimates simpler, we will hereafter {adopt} a hybrid
framework in which we use the simpler Newtonian expressions for the drag
forces but evaluate the mass and momentum fluxes needed for these
expressions within in a full general relativistic prescription. Hence, we
write the \textit{total} drag force in the $i$th direction as
\begin{eqnarray}
  F^{i}= F^{i}_{\rm drag} + F^{i}_{\rm dyn}\,,
  \label{eq:totalforce}
\end{eqnarray}
where the first contribution is due to the accretion of the linear
momentum and we compute it as
\begin{eqnarray}
F^{i}_{\rm drag} =\frac{1}{\Sigma} \int \alpha \sqrt{\gamma} \big(
\dot{\boldsymbol{P}}
\cdot \boldsymbol{x}\big)^{i} d\Sigma\,,
\label{eq:dragi}
\end{eqnarray}
where $\Sigma$ is a spherical surface at a given distance from the black
hole, $\boldsymbol{e}_x$ is the unit vector in the direction of motion of
the black hole (the negative $x$-direction in our setup), and
$\dot{P}^{i}$ is computed using Eq. (\ref{eq:pdot}). The second
contribution is instead due to dynamical friction arising from the
interaction of the black hole with the matter of the common envelope
\citep{Ostriker1999, Barausse2007, Barausse2007b}.

Starting from the Newtonian approximation to the infinitesimal dynamical
friction force $dF_{\rm dyn}$ produced by fluid element of density $\rho$
in a volume $dV$, \ie $d\boldsymbol{F}_{\rm dyn} = M \rho\, dV
\boldsymbol{r}/r^{2}$ (see Equation (26) of \cite{MacLeod2017}), we express
the relativistic dynamical friction as
\begin{eqnarray}
F^{i}_{\rm dyn} = M \int \alpha \sqrt{\gamma} D
\left(\frac{\boldsymbol{e}_{r}\cdot \boldsymbol{e}_{x}}{r^{2}}\right)^{i}
d^{3}V\,,
\label{eq:dyn}
\end{eqnarray}
where the volume integral in Eq. (\ref{eq:dyn}) is computed from the
event horizon up to a given radius where the rate is extracted (see Fig.
\ref{fig:dragforcedetectors}), and $\boldsymbol{e}_{r}$ is the unit
vector in the radial direction.

In Fig. \ref{fig:dragforcedetectors} we show the evolution of the total
drag force \eqref{eq:totalforce} as the simulation progresses and when
measured at detectors placed at different spherical radii normalized to
the accretion radius. {The top} panel shows the total drag force for a
dimensionless density gradient $\epsilon_{\rho}=0.5$, while the {bottom}
panel is for $\epsilon_{\rho}=1.0$; both panels refer to $\Gamma=4/3$. The
time is scaled by the crossing time, and to help in the comparison with
the Newtonian results in \citet{Murguia-Berthier2017845}, we normalize
with respect to the reference relativistic drag rate given by Equation 
\eqref{eq:canpdot}.

In analogy with the Newtonian results, Fig. \ref{fig:dragforcedetectors}
shows that the drag increases when measured at increasingly large radii
and that it changes sign, becoming positive when measured outside a
surface of radius $r \sim 0.28 r_{\rm acc}$. Note the very smooth
behavior of the drag as a function of time, which is in contrast with
the stochastic behavior seen in the Newtonian simulations of
\cite{MacLeod2017}, and may be due to the turbulent nature of the flow
near the sink.

\begin{figure*}
  \center
  \includegraphics[width=0.75\columnwidth]{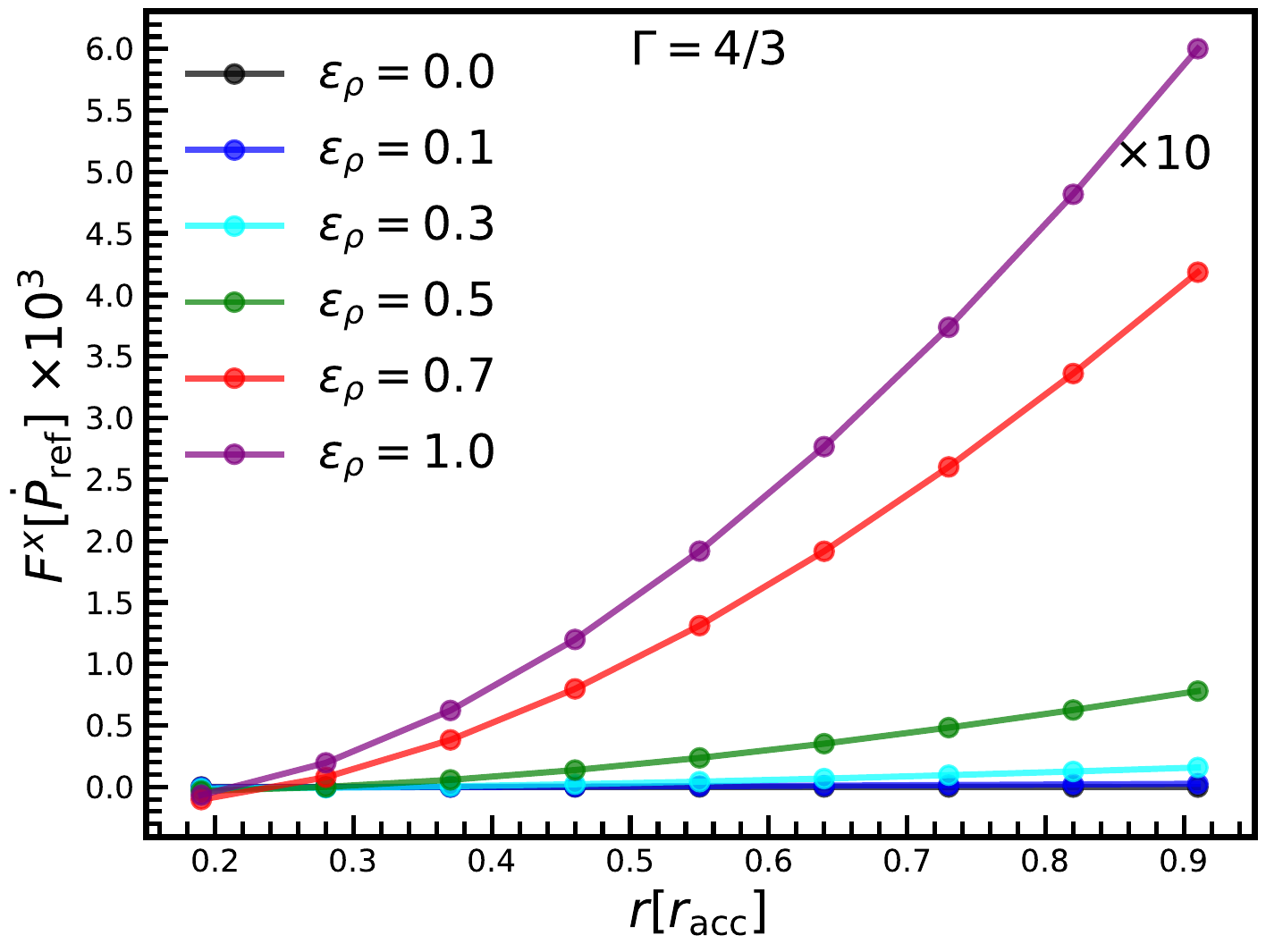}
  \hskip 1.0cm
  \includegraphics[width=0.75\columnwidth]{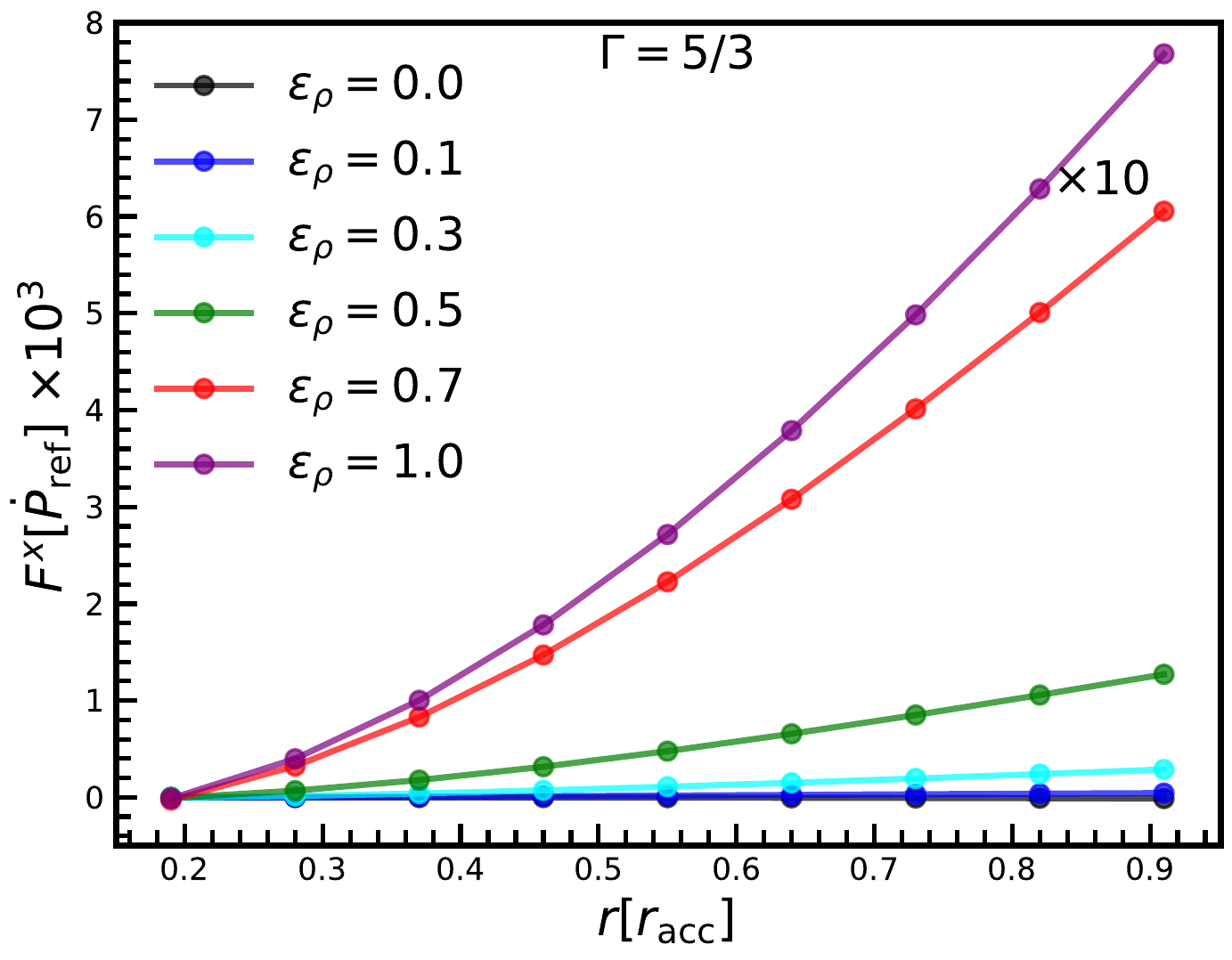}
  \caption{Total drag forces as measured at different spherical surfaces
    expressed in terms of the accretion radius (see Equation (\ref{eq:totalforce})) 
    and computed after the evolution has
    reached a stationary state, \ie for $t>20 t_{\rm cross}$. The left
    and right panels refer to $\Gamma=4/3$ and $\Gamma=5/3$,
    respectively.}
    \label{fig:forceversura}
\end{figure*}

{Figure \ref{fig:forceversura} summarizes} the results for the total drag
force -- once it has reached a steady-state value -- and reported as a
function of the detector position (for $r > 0.28\,r_{\rm acc}$), as well
as for the various density gradients and adiabatic indices (left and
right panels for $\Gamma=4/3$ and $\Gamma=5/3$, respectively). Note that
the total drag force increases monotonically for $r_{_{\rm EH}}< r <
r_{\rm acc}$ (changing sign at $\sim 0.28\, r_{\rm acc}$) and that these
increases are larger for more severe density gradients in the stellar
envelope.

The simple dependence of the total drag force from the density gradient
allows us to describe all of the simulation data in terms of a simple
phenomenological expression,
\begin{eqnarray}
  F^{x}(r,\epsilon_{\rho}) = \dot{P}_{\rm ref}\left(  \omega_{1}(r) +
  \omega_{2}(r)\epsilon_{\rho} +
  \omega_{2}(r)\epsilon^{2}_{\rho}\right)\,,
  \label{eq:drag}
\end{eqnarray}
where $\dot{P}_{\rm ref}$ is the reference linear momentum rate given in
Eq. (\ref{eq:canpdot}) and the coefficients $\omega_{i}(r)$ are reported
in {Table \ref{tab:forcecoef}} for $r=r_{_{\rm EH}}$ and $r=r_{\rm
  acc}$. Note that although the Newtonian total drag force has the same
functional dependence as in expression \eqref{eq:drag}, the values of the
coefficients are rather different \citep{MacLeod2015}.

\begin{figure*}
  \center
  \includegraphics[width=0.31\textwidth]{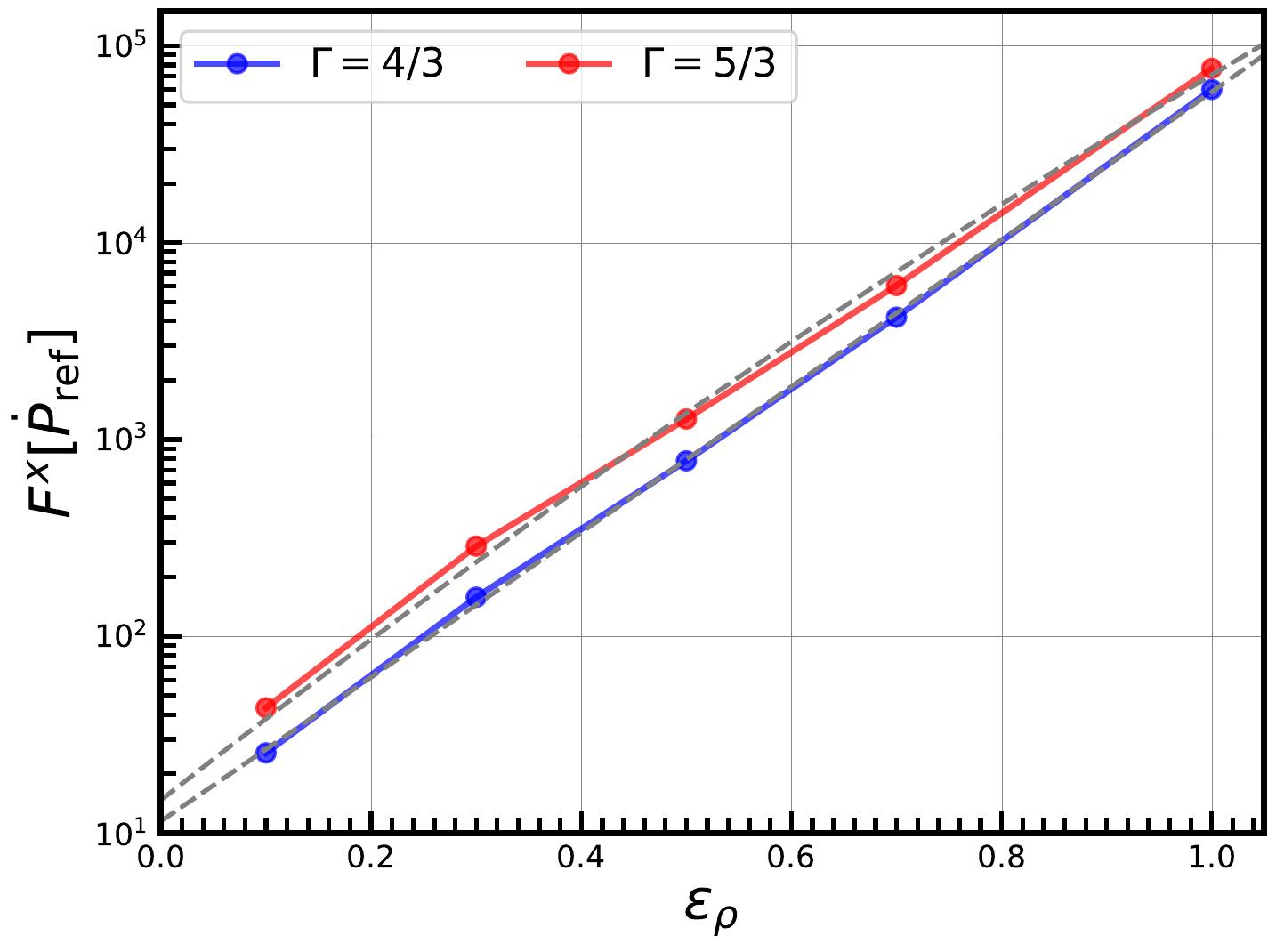}
  \hskip 0.15cm
  \includegraphics[width=0.31\textwidth]{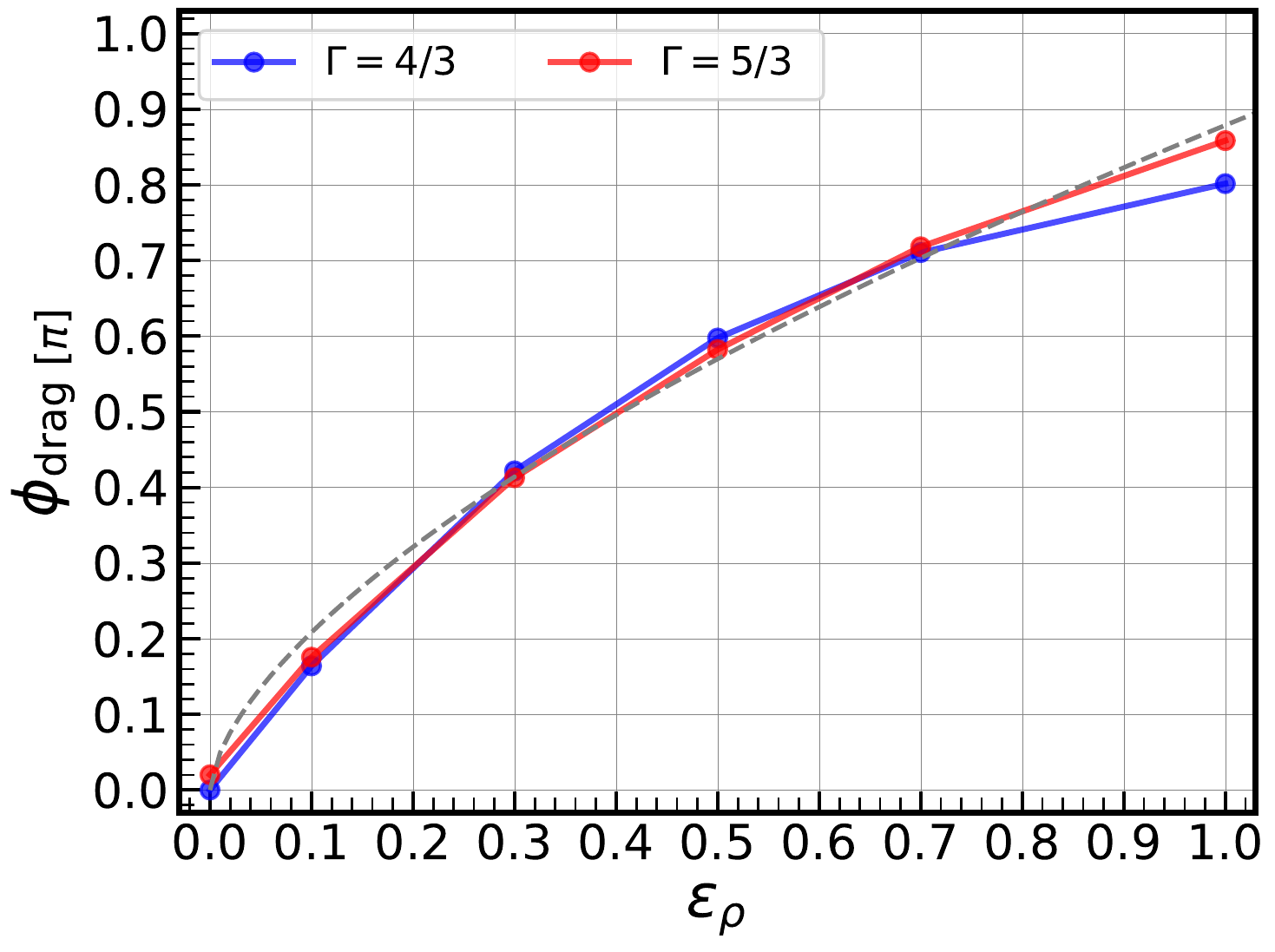}
  \hskip 0.15cm
  \includegraphics[width=0.35\textwidth]{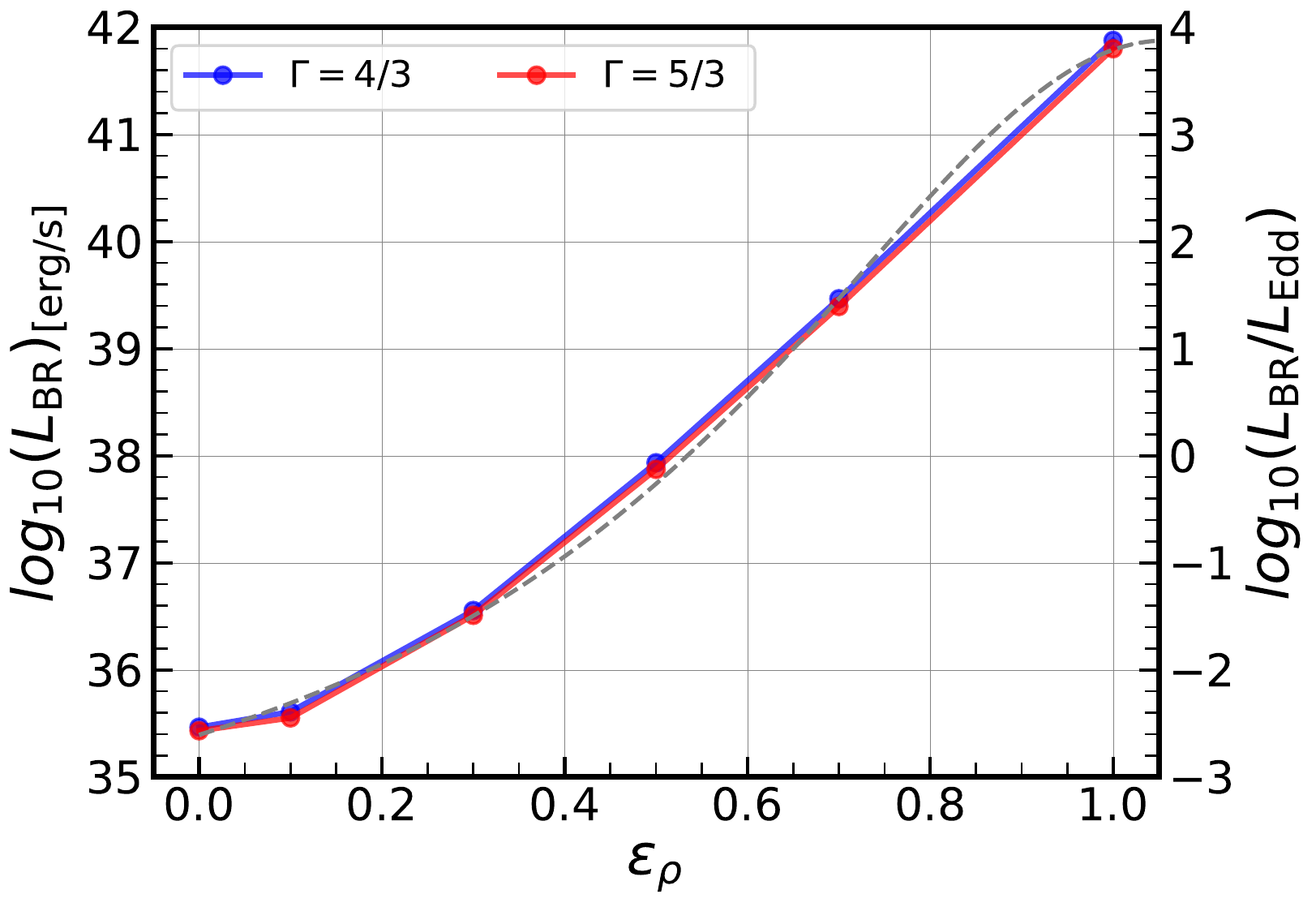}
  \caption{Left panel: Drag force as a function of the
    dimensionless density gradient. All positive forces measured at
    accretion radius $r_{\rm acc}$ are plotted with corresponding fit
    function given in Eq. ( \ref{eq:drag} ). Middle panel: Drag
    angle between $x$-axis and the shock-cone axis as produced by the
    density gradients in the envelope. This angle is related to the
    deviation of the trajectories of the black hole, modifying the orbit
    of the binary system. Right panel: bremsstrahlung luminosity
    from interaction of the black hole with a supergiant-star envelope
    expressed both in $\rm{erg\, s^{-1}}$ and in terms of the Eddington
    luminosity $L_{\rm Edd}$.}
    \label{fig:force}
      \vskip 0.5cm
\end{figure*}

The left panel of the Fig.\ref{fig:force} reports the function $F^{x}$ at
the event horizon as function of the density gradient and for the two
adiabatic indices, with the dotted lines referring to the analytic
expression \eqref{eq:drag}.

\begin{table} 
\begin{center}
\caption{Dimensionless units coefficients $\omega_{i}$ total force in the
  $x$-direction, direction of the black hole's motion given by equation
  (\ref{eq:drag}) measured at event horizon and at the accretion radius
  $r_{\rm acc}$.  }
\begin{tabular}{lcccccc}
\hline\hline
$\Gamma$& $\omega_{1}$&$\omega_{2}$ &     $\omega_{3}$  &    $\ell_{1}$    &     $\ell_{2}$    &    $\ell_{3}$  \\
\hline
   &      &  $r\sim r_{_{\rm EH}}$    &   && &\\
\hline
$4/3$      &    $0.54 $   &  $5.79$    &  $1.02$ & $0.542$  &  $-0.798$  & $0.377$  \\
\hline
$5/3$      &     $0.83$  &  $5.46$    & $1.36$   & $0.553$  &  $-0.819$  & $0.387$  \\
\hline
           &             &  $r\sim r_{\rm acc}$ &  &&&\\
\hline
$4/3$      &    $2.44$   &  $8.39$    & $0.15$   & $-$ & $-$ & $-$ \\
\hline
$5/3$      &     $2.69$  &  $9.62$    & $-1.14$  & $-$ & $-$ & $-$ \\
\hline
\end{tabular}
\label{tab:forcecoef}
\end{center}
\end{table}

As mentioned in Sec. \ref{ss:Morpho} and as shown in the schematic
diagram shown in Figure \ref{fig:tzo}, the presence of a density gradient leads to a
deviation of the axis of the shock cone away from asymptotic direction of
motion of the fluid. Furthermore, the angle measuring this deviation
increases nonlinearly with the density gradient, becoming even larger
than $\pi/2$ for $\epsilon_{\rho}=1$. Measuring this angle is important,
as it can be used to quantify the deviation of the black hole's orbit
away from a quasi-circular orbit.

Using the data in our simulations once the matter flow has reached a
stationary configuration, we can measure this drag angle $\phi_{\rm
  drag}$ and we have reported the corresponding values as a function of
$\epsilon_{\rho}$ in the middle panel of Fig. \ref{fig:force}. The simple
functional behavior allows us to express the results of the numerical
simulations in terms of the simple fitting function (shown as a dotted line
in the middle panel of Fig. \ref{fig:force})
\begin{eqnarray}
  \label{eq:theta_drag}
  \phi_{\rm drag}= {\cal Q}\, \epsilon^{5/8}_{\rho}\,,
\end{eqnarray}
where the constant ${\cal Q}=2.761$ for all models considered
here. Interestingly, expression \eqref{eq:theta_drag} appears
``universal'' in the sense that it depends only very weakly on the
adiabatic index of the envelope.

We conclude this Section by considering what could be the radiative
signatures of the accretion processes considered here. While we do not
have the ambition of performing accurate general relativistic,
radiative transfer calculations as those performed by
\citep{Zanotti2011,Roedig2012} or \citep{Fragile2014}, it is useful to
provide here a rough estimate of the luminosity produced considering
bremsstrahlung processes from electron--proton collisions
\citep{Rybicki_Lightman1986,Zanotti2010}
\begin{eqnarray}
L_{_{\rm BR}}&=& 3.0 \times 10^{78}  \int \left( T^{\frac{1}{2}} \rho^{2}
\sqrt{\gamma} dV \right) \left( \frac{M_{\odot}}{M} \right) \ {\rm  \frac{erg}{s}}\,,
\label{eq:luminosity}
\end{eqnarray} 
which can be readily computed in terms of the quantities evolved in the
simulations.

The right panel of Fig. \ref{fig:force} summarizes the results of the
simulations as a function of the density gradient and shows that -- as
expected -- the bremsstrahlung luminosity grows with the density gradient
and almost exponentially for large gradients. The simple functional
behavior of $L_{_{\rm BR}}$, which is also essentially independent of
the adiabatic index, allows for a simple fitting expression of the type
\begin{eqnarray}
  \log \left(\frac{L_{_{\rm BR}}}{L_{\odot}}\right) = \frac{1}
       {\ell_{1}+\ell_{2}\epsilon_{\rho} + \ell_{3}\epsilon^{2}_{\rho}}\,,
\label{eq:fitluminosity}
\end{eqnarray}
where the values of $\ell_{i}$ can be found in Table \ref{tab:forcecoef}
and the analytic behavior is shown as a dotted line in the right panel
of Fig. \ref{fig:force}.
\section{Conclusions}
\label{sec:Sum}

Using general relativistic simulations, we have carried out a systematic
investigation of the properties of the accretion flow onto a nonrotating
black hole moving supersonically in a medium with regular but different
density gradients. This scenario has been here considered with the goal
of providing more accurate and realistic estimates of the secular
behavior of the mass accretion and drag rates in the ``common-envelope''
scenario encountered when a compact object -- a black hole or a neutron
star -- moves in the stellar envelope of a red supergiant star.

The simulations reveal that the supersonic motion always rapidly reaches
a stationary state and it produces a shock cone in the downstream part of
the flow. At the same time, the gradual change in the density gradient
leads to a smooth change general behavior of the flow and in all of the
relevant accretion rates. In particular, in the absence of density
gradients (\ie $\epsilon_{\rho}=0$), we recover the phenomenology already
observed in the well-known Bondi--Hoyle--Lyttleton accretion problem, with
a shock cone and whose axis is stably aligned with the direction of
motion \citep{Font1999b, Donmez2010, Cruz2012}. However, as the density
gradient is increased and the black hole encounters a nonuniform
medium, the system {reaches a steady state where the flow and its
  discontinuities have found a new equilibrium. When such a state is
  found -- which requires longer times for large density gradients --}
the shock cone is progressively and stably dragged toward the direction
of motion. With sufficiently large gradients, the shock-cone axis can
become orthogonal to the direction, or even move in the upstream region
of the flow in the case of the largest density gradient (\ie
$\epsilon_{\rho}=1$). In all cases considered -- which refer to fluids
with adiabatic index set to $\Gamma=4/3$ or $5/3$ -- the shock cone heats
up the fluid, increasing the temperature by about one order of magnitude
with respect to the rest of the envelope, and of almost two orders of
magnitude hotter than in the initial state. The fluid is also compressed
by the shock, reaching rest-mass densities that are about two orders of
magnitude larger than in the unperturbed fluid.

Together with the phenomenological aspects of the accretion flow, we have
also investigated and quantified the rates of accretion of mass and
momentum onto the black hole, which are particularly important 
because they ultimately play a leading role in the secular evolution and modeling of
the common-envelope scenario. More specifically, and as may have been
expected, the rates of accretion of mass and momentum (either radial or
angular) increase with the density gradient. Furthermore, this increase
becomes essentially exponential as soon as the dimensionless density
gradient is $\epsilon_{\rho}\gtrsim 0.2$, thus allowing us to derive
simple phenomenological expressions for the mass accretion rate and the
momentum accretion rate. Equally simple analytic expressions have been
found for the total drag force -- which is a combination of the drag due to
due the accretion of linear momentum and of the drag due to dynamical
friction -- of the drag angle of the shock cone and of the
bremsstrahlung luminosity that has been estimated. All of these
expressions can be readily employed in the astrophysical modeling of the
long-term evolution of a binary system experiencing a common-envelope
evolution. Furthermore, since the mass accretion rates measured in
nonuniform media are well above the Eddington limit, this increases the
chances of observing this process also in binary systems of stellar-mass
black holes \citep{Caputo2020}.

Finally, we have carried out a systematic comparison with previous
Newtonian results, finding that while the general relativistic
phenomenology and rates are comparable in the case of the
Bondi--Hoyle--Lyttleton scenario (\ie for $\epsilon_{\rho}=0$), significant
differences develop for nonzero density gradients. More specifically,
while in a general relativistic context the rates increase with the
density gradient, the opposite is true for calculations performed in a
Newtonian framework \citep{MacLeod2015}. The robustness of the increase
of the rates with $\epsilon_{\rho}$ within a relativistic framework has
been validated by performing the same simulations with an independent
general relativistic MHD code, obtaining results that differ by less than
$5\%$.

This different behavior may be due to the different relative strength of
gravitational forces in the two contexts, but it may also be due to the
different boundary conditions, which we impose inside the event horizon
by using the ingoing Eddington--Finkelstein coordinates and are those of a
purely infalling gas. On the other hand, the Newtonian framework requires
the introduction of a ``sink'' surrounding the central point mass and
whose gravitational potential is smoothed within this sink. The different
physical conditions near the black hole (or sink in the Newtonian
framework) may also be responsible for the lack of a turbulent flow and
vorticity in our simulations, which instead appear in the Newtonian
simulations.

The results presented here can be improved in a number of ways. First, we
can consider more realistic black holes with nonzero spin; while the spin
of black holes in such binary system is currently unknown, determining
the dependence of the accretion rates on the black hole spin is of great
importance. Second, we can include the contribution of a radiation field --
which is expected to decrease but not suppress the mass accretion
rate\footnote{{The flow conditions in \citet{Zanotti2011} are somewhat
    milder than those considered here, \ie $v_{\infty} = 0.1,
    c_{s,\infty}=0.07, \mathcal{M}_{\infty}=1.4$, yet radiation pressure
    was not sufficient to quench accretion. This leads to our expectation
    that this may happen also for the flow conditions considered here,
    \ie $v_{\infty} = 0.2, c_{s,\infty}=0.1,
    \mathcal{M}_{\infty}=2.0$. However, proper self-consistent
    simulations are needed to assess whether this expectation is
    correct.}}  \citep{Zanotti2011} -- by solving, together with the
relativistic hydrodynamic equations, also those of relativistic radiative
transfer. Finally, we can account for the presence of magnetic fields,
which may alter the dynamics of the flow and hence the various accretion
rates. We plan to address these additional aspects of the
general relativistic common-envelope problem in future works.

\section*{Acknowledgements}

{We thank Morgan MacLeod and Enrico Ramirez-Ruiz for useful feedback, and
  we are grateful to the anonymous referee for the constructive
  suggestions.}  Support comes also in part from "PHAROS'', COST Action
CA16214; LOEWE-Program in HIC for FAIR; European Union's Horizon 2020
Research and Innovation Programme (Grant 671698) (call FETHPC-1-2014,
project ExaHyPE); and the ERC Synergy Grant ``BlackHoleCam: Imaging the Event
Horizon of Black Holes'' (grant No. 610058). The simulations were
performed on the SuperMUC cluster at the LRZ in Garching, on the LOEWE
cluster in CSC in Frankfurt, and on the HazelHen cluster at the HLRS in
Stuttgart.

\appendix

\section{A. Additional astrophysical considerations}
\label{sec:appA}

\subsection{A.1. Inspiral timescales and different initial data}

Following \cite{MacLeod2015b}, in this appendix we estimate the inspiral
timescale as a function of the density gradients $\epsilon_{\rho}$ and of
the accretion rates measured in our simulations. In particular, we define
the inspiral timescale as \citep{MacLeod2015b}
\begin{equation}
  \label{eq:tau_insp}
  \tau_{\rm insp}:= \frac{E_{\rm orb}}{\dot{E}_{\rm acc}}\,,
\end{equation}
where $E_{\rm orb}$ is the orbital energy that, for a binary system of
masses $M$ and $M_{\rm star}$, and separation $a$, is given by (for
clarity, we restore the use of the gravitational constant and of the
speed of light)
\begin{equation}
  E_{\rm orb}:=\frac{M\,M_{\rm star}}{2a}\,.
\end{equation}
On the other hand, $\dot{E}_{\rm acc}$ is the rate of dissipation of the
kinetic energy via accretion, which we express in terms of the measured
accretion rates and asymptotic velocity as
\begin{equation}
  \dot{E}_{\rm acc}=\dot{M} v_{\infty}^{2}\,,
\end{equation}
where the accretion rates are those reported in
Fig. \ref{fig:ratesvstime}. 

\begin{figure}[!ht]
  \center
  \includegraphics[width=0.4\textwidth]{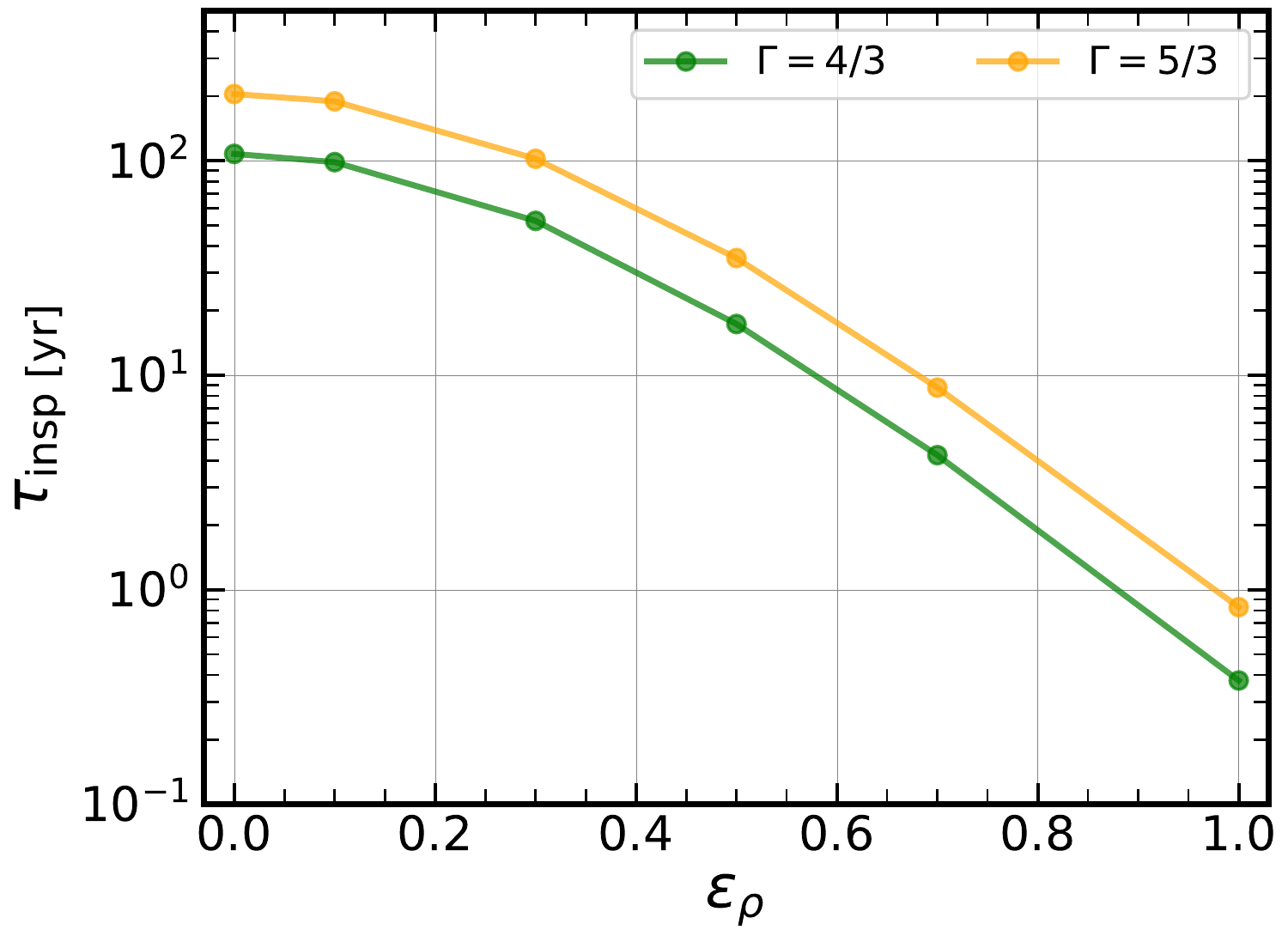}
  \caption{Inspiral timescale (see Equation \eqref{eq:tau_insp}) as function
    of the dimensionless density gradients $\epsilon_{\rho}$ and the two
    equations of state considered. The initial separation is assumed to be $a =R_{\rm
      star}$ and the mass accretion rates are those reported in
    Fig. \ref{fig:ratesvstime}.}
    \label{fig:tinspiral}
\end{figure}

To fix numbers, we evaluate the inspiral timescale after setting $M_{\rm
  star}=16 M_{\odot}$ and $M=4 M_{\odot}$ and, as in the rest of the
paper, $v_{\infty}=0.2\,c$. The results are shown in
Fig. \ref{fig:tinspiral}, which reports the inspiral timescale for an
initial separation of $a=R_{\rm star}$, \ie when the black hole is just
on the outer parts of the common envelope, thus representing the upper
limit for the timescale. Under these conditions, and using the accretion
rates reported in Fig. \ref{fig:ratesvstime}, we found that the
$\tau_{\rm insp}$ $\sim 100\,( 200) {\rm yr}$ for $\Gamma=4/3\, (5/3)$
and for motion in a uniform medium ($\epsilon_{\rho}=0$). Such timescale
decreases by about two orders of magnitude when the considering larger
gradients in density. In the extreme case of
$\epsilon_{\rho}=1$, the inspiral timescale is as short as $\sim 0.4\,
(0.8)\, {\rm yr}$.

\subsection{A.2. Impact of the initial data}
\label{sec:iotid}

It is clear that the precise values of the accretion rates
  ultimately depend on the initial conditions considered, since at the
  end what matters in determining the mass accretion rate is the size of
  the mass current near the black hole horizon. In order to investigate a
  different set of initial conditions, we have carried out simulations
  employing the initial conditions used in \citet{Zanotti2011}, \ie
  $v_{\infty} = 0.1$ and $c_{s,\infty}=0.07$, thus yielding a Newtonian
  Mach number $\mathcal{M}_{\infty} \simeq 1.4$. Such a configuration
  should be contrasted with the reference configuration discussed in the
  main text, for which $v_{\infty} = 0.2$, $c_{s,\infty}=0.1$, and thus
  $\mathcal{M}_{\infty}=2.0$. In essence, the initial conditions in
  \citet{Zanotti2011} refer to slower and more tenuous fluid, but also to
  a flow with a larger accretion radius, \ie $r_{\rm acc} \simeq 67\,M$
  instead of $r_{\rm acc} \simeq 20\,M$. As a result, the computational
  domain -- whose radial extent is set to be $r_{\rm max}=10\,r_{\rm
    acc}$ -- needs to be increased by almost of a factor of three, with a
  consequent increase in the computational costs.

\begin{figure}[!ht]
  \center
  \includegraphics[width=0.4\textwidth]{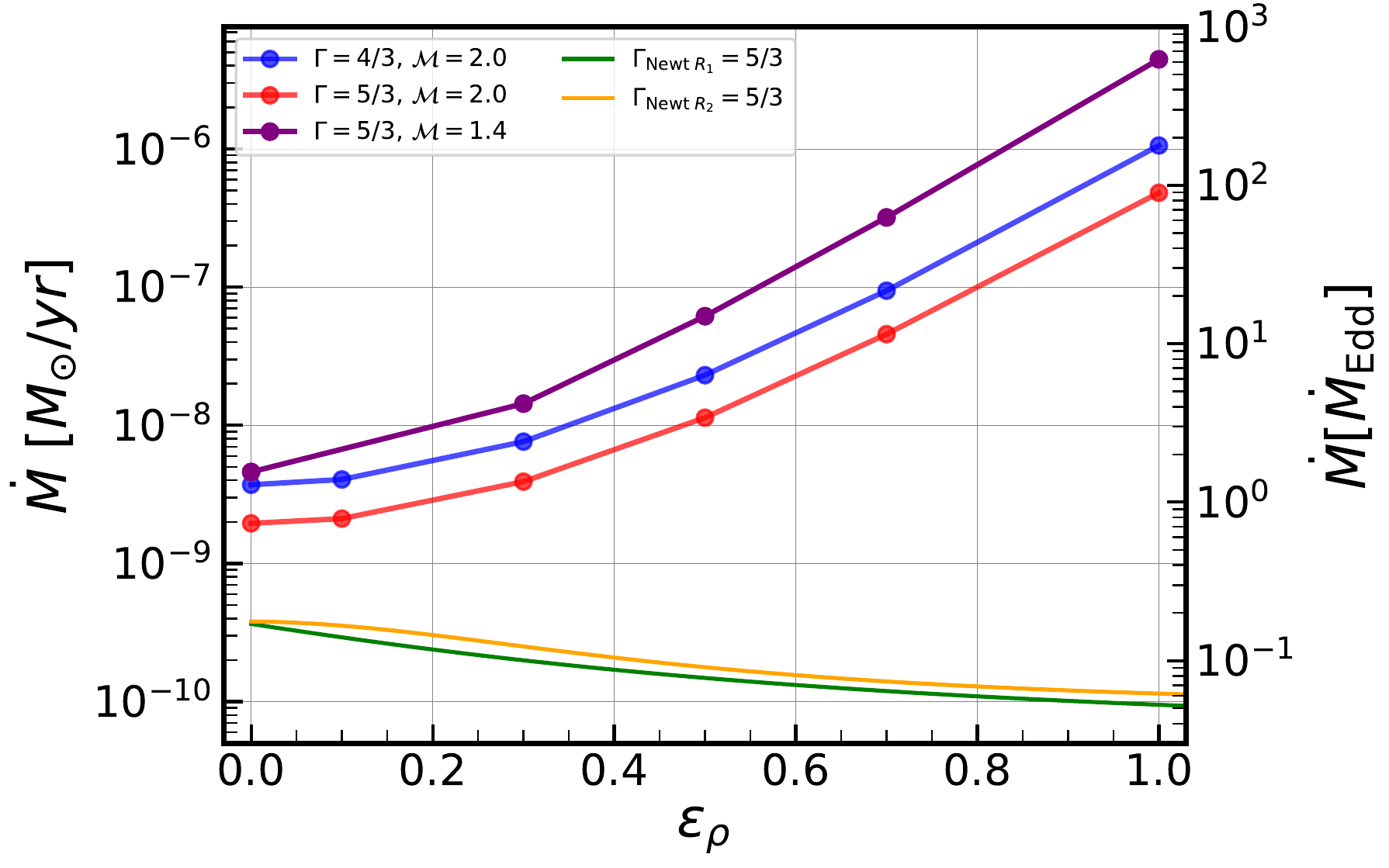}
  \caption{Same as Fig. \ref{fig:rates} but also showing the accretion
    rate when considering the initial flow discussed in
    \citet{Zanotti2010} (purple line). Note that despite that fluid is in
    this case less dense and with a smaller initial velocity, the
    accretion rate is actually larger because the fluid can be compressed
    more and this leads to an increase in the mass current near the black
    hole.}
    \label{fig:rates3}
\end{figure}

 The results of this investigation for five different values of
   the dimensionless density gradient are reported in
   Fig. \ref{fig:rates3}, which is very similar to Fig. \ref{fig:rates},
   but where we also show the accretion rate when considering the initial
   flow discussed in \citet{Zanotti2011} (purple line) for an adiabatic
   index $\Gamma=5/3$. Note that despite that fluid is in this case less
   dense and with a smaller initial velocity, the accretion rate is
   actually larger because the fluid can be compressed more and this
   leads to an increase in the mass current near the black hole. While
   these results highlight that the mass accretion rates also depend on
   the initial conditions considered, it is useful to note that the
   reference fluid conditions adopted in the main text (\ie
   $\mathcal{M}_{\infty}=2.0$) provide mass accretion rates smaller than
   those that would be obtained if a more tenuous gas in the stellar
   envelope were considered. In turn, because smaller values of
   $\mathcal{M}_{\infty}$ yield smaller accretion rates, it will
   be easier to reduce them when switching on the coupling with a
   radiation fluid as done in \citet{Zanotti2011}.  A more systematic
   analysis of the dependence of $\dot{M}$ on $\rho_{\infty}$ and
   $\mathcal{M}_{\infty}$ will be part of our future work.

\section{B. Additional numerical details}
\label{sec:appB}

\subsection{B.1. Convergence test}

The numerical results presented in this work are solving consistently the
accretion radius by $200$ cell grids (see Tab. \ref{tab:params}),
however, in this appendix we also show the convergence test for the model
\texttt{RCE.1.0.5o3} that corresponds to the higher density gradient
studied here. For this test, three simulations have been performed using
resolutions of $1000 \times 128$, $2000 \times 256$ (which is also the
resolution used for all of the results presented so far) and $4000 \times
512$ zones in a uniform grid. Hereafter, we will refer to these
resolutions as the low (L), medium (M), and high (H) resolutions.

\begin{figure}[!ht]
  \center
  \includegraphics[width=0.37\textwidth]{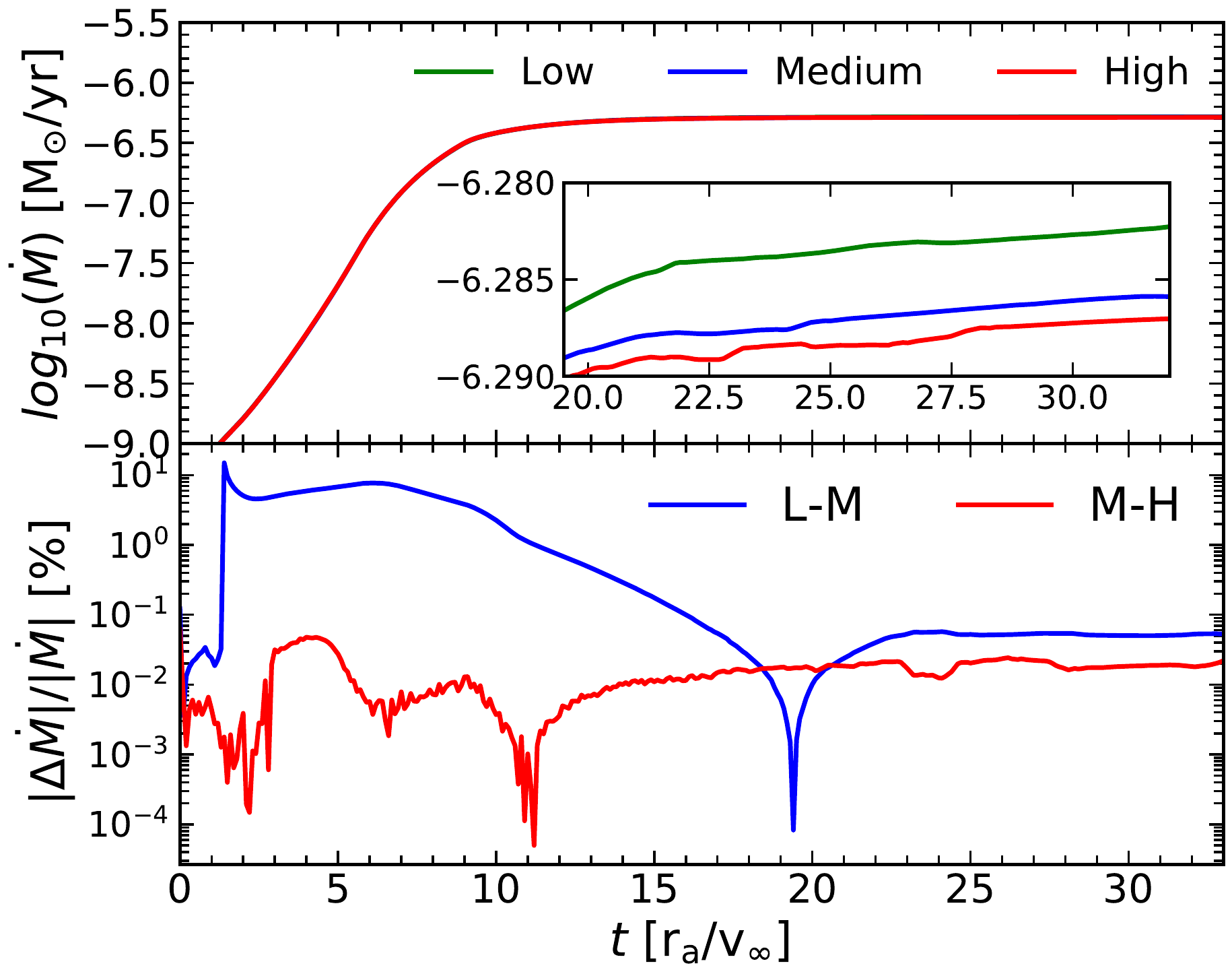}
  \hskip 1.0cm
  \includegraphics[width=0.4\textwidth]{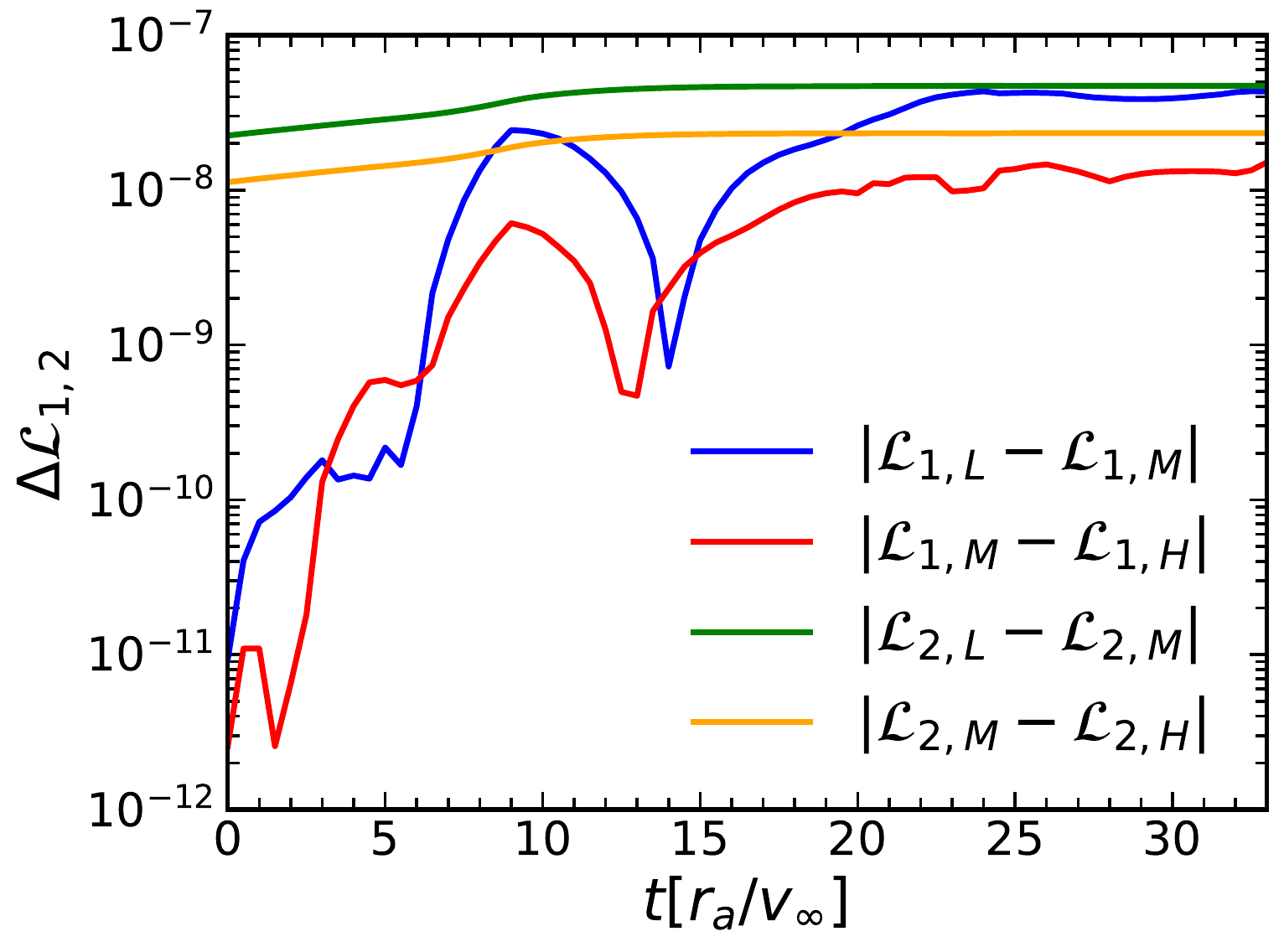}
  \caption{Left panel: mass accretion rates at the L, M, and H
    resolutions (top) and the relative differences (bottom) for model
    \texttt{RCE.1.0.5o3}. The differences between the high resolution and
    medium (our simulation) are less than $0.1\%$ when the gas reaches to
    the steady state, as well as when comparing the low and medium
    resolutions. Right panel: differences in the
    $\mathcal{L}_{1}$ and $\mathcal{L}_{2}$ norms of the rest-mass
    density at the three different resolutions.}
    \label{fig:conv}
\end{figure}

In particular, we show in the left panel of Fig. \ref{fig:conv} the
mass accretion rates at different resolutions (top) and the relative
differences (bottom) for model \texttt{RCE.1.0.5o3}. Note that the when
the flow reaches a steady state, \ie after $\sim 20$ crossing times, the
relative differences are less than $0.1\%$, between both the $\rm H/M$
resolutions and the $\rm M/L$ resolutions. These small differences clearly
indicate that the asymptotic rates computed with the M resolution are
robust and accurate. The right panel of Fig. \ref{fig:conv}, on the other
hand, reports the differences in the $\mathcal{L}_{1}$ and
$\mathcal{L}_{2}$ norms of the rest-mass density at different
resolutions, \eg $\mathcal{L}_{1,L} - \mathcal{L}_{1,M}$, as a function
of time. Note that the differences that the differences in the
$\mathcal{L}_{1,2}$ norms are of the order of $\sim 10^{-8}$ and $\sim 5
\times 10^{-8}$, respectively. Furthermore, using these norms, it is
possible to validate that the code is in a convergence regime with
convergence order between $1.2$ and $2.0$ (not shown in
Fig. \ref{fig:conv}).

\subsection{B.2. Comparison between \texttt{BHAC} and \texttt{CAFE}}
\label{sec:appC}

As mentioned in the main text, the numerical results reported in this
paper have been produced making use of the \texttt{CAFE} code
\citep{Lora2015}. At the same time, and for the purpose of validating the
accuracy of the results -- some of which are considerably different from
the Newtonian ones -- about half of the simulations have also been
reproduced by a more recent and advanced numerical code solving the
equations of general relativistic MHD: \texttt{BHAC}
\citep{Porth2017}. More specifically, all models with adiabatic index
$\Gamma=5/3$ have been simulated by both codes for the six values of the
density gradient parameter (see Table \ref{tab:params}) after using the same
numerical methods, grid resolution, and computation domain as described
in Section \ref{sec:Met}.

\begin{figure}[!ht]
  \center
  \includegraphics[width=0.4\textwidth]{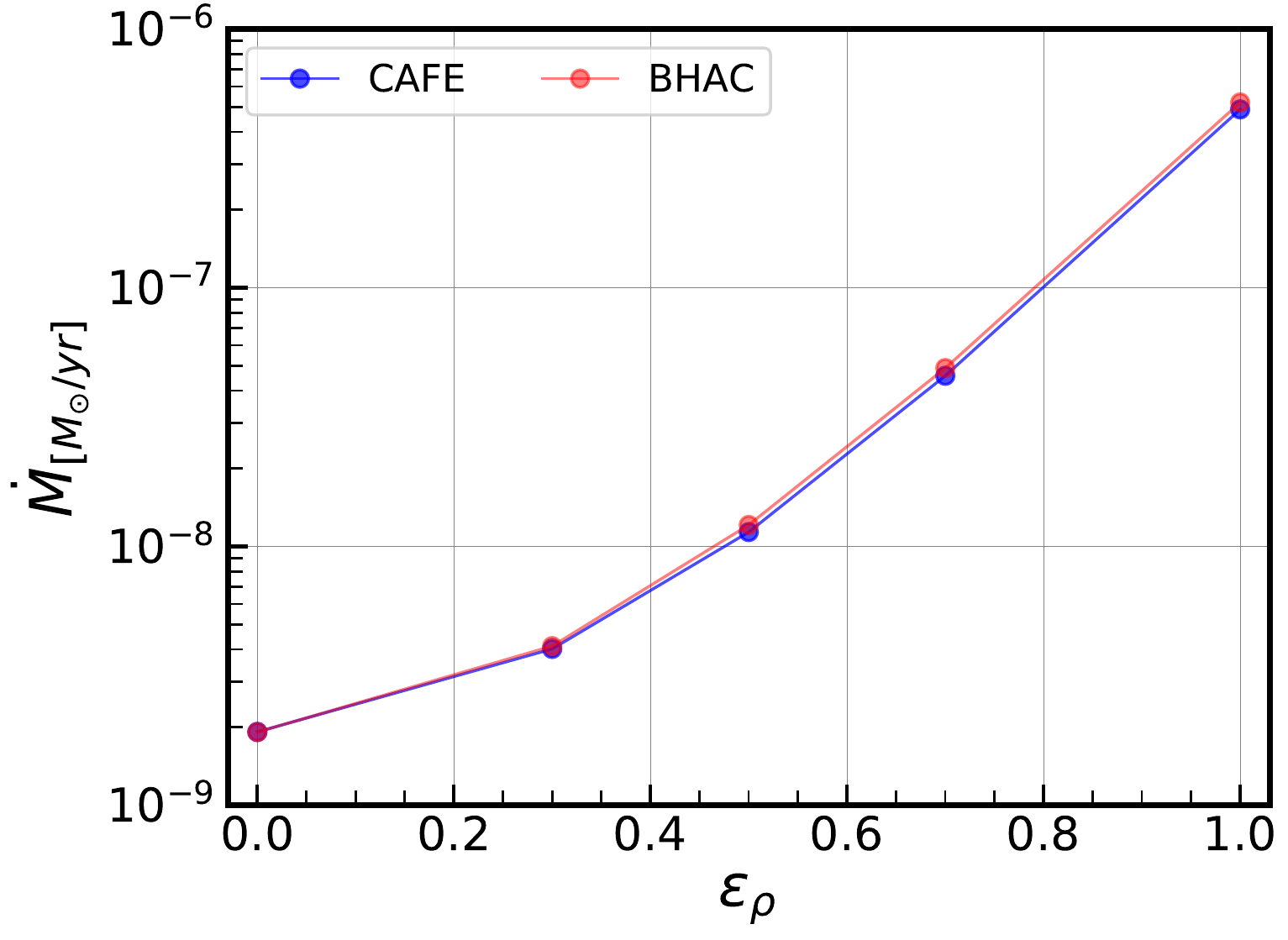}
  \caption{mass accretion rate as a function of the density gradient when
    computed by \texttt{CAFE} (blue line) or by \texttt{BHAC} (red line)
    for models $\texttt{RCE.0.0.5o3}-\texttt{RCE.1.0.5o3}$. In both
    cases, the mass accretion rate is measured at the event horizon and
    when a stationary regime in the accretion has been reached, \ie at $t
    \sim 5000\,M$. The relative differences between the two codes are
    below $5\%$ and both codes report an increase of the mass accretion
    rate with the density gradient.}
    \label{fig:bhac}
\end{figure}

The results of a comparison between the two codes are shown in
Fig. \ref{fig:bhac} for what is arguably the most important quantity
among the ones measured: the value of the stationary mass accretion rate
as a function of the density parameter. This is shown in
Fig. \ref{fig:bhac}, where we show the mass accretion rate in solar
masses per year as computed by \texttt{CAFE} (blue line) or by
\texttt{BHAC} (red line) for models
$\texttt{RCE.0.0.5o3}-\texttt{RCE.1.0.5o3}$. In both cases the
mass accretion rate is measured at the event horizon and when a
stationary regime in the accretion has been reached, \ie at $t \sim
5000\,M$. Despite the numerous differences between the two codes, the
rates computed are very similar, with differences that are below
$5\%$. More importantly, both codes show that the mass accretion rate
increases with the density gradient (see Equation \eqref{eq:fitmom}),
confirming that this is the correct behavior and that the decrease
measured in Newtonian calculations is due either to the different
strength of the gravitational field or to the different boundary
conditions applied at the surface of the compact object.

\end{document}